\documentclass[fleqn,usenatbib]{mnras}

\usepackage{newtxtext,newtxmath}

\usepackage[T1]{fontenc}

\DeclareRobustCommand{\VAN}[3]{#2}
\let\VANthebibliography\thebibliography
\def\thebibliography{\DeclareRobustCommand{\VAN}[3]{##3}\VANthebibliography}

\usepackage{graphicx}	
\usepackage{amsmath}	

\title[Mrk\,231]{AGN Jets and Winds in Polarised Light: The Case of Mrk\,231}

\author[Silpa et al.]{
Silpa S.,$^{1}$\thanks{E-mail: silpa@ncra.tifr.res.in (SS)}
P. Kharb,$^{1}$
C. P. O' Dea,$^{2}$
S. A. Baum,$^{2}$
B. Sebastian,$^{3}$
D. Mukherjee,$^{4}$
C. M. Harrison$^{5}$
\\
$^{1}$National Centre for Radio Astrophysics (NCRA) - Tata Institute of Fundamental Research (TIFR), S. P. Pune University Campus, Ganeshkhind, Pune 411007, India\\
$^{2}$Department of Physics and Astronomy, University of Manitoba, Winnipeg, MB R3T 2N2, Canada\\
$^{3}$Department of Physics and Astronomy, Purdue University, 525 Northwestern Avenue, West Lafayette, IN 47907, USA\\
$^{4}$IUCAA, S. P. Pune University Campus, Ganeshkhind, Pune 411007, India\\
$^{5}$School of Mathematics, Statistics and Physics, Newcastle University, Newcastle upon Tyne, NE1 7RU, UK\\
}

\pubyear{2021}

\begin{document}
\label{firstpage}
\pagerange{\pageref{firstpage}--\pageref{lastpage}}
\maketitle

\begin{abstract}
We present the results of a multi-frequency, multi-scale radio polarimetric study with the Very Large Array (VLA) of the Seyfert 1 galaxy and BALQSO, Mrk\,231. We detect complex total and polarized intensity features in the source. Overall, the images indicate the presence of a broad, one-sided, curved outflow towards the south which consists of a weakly collimated jet with poloidal inferred magnetic fields, inside a broader magnetized ``wind'' or ``sheath'' component with toroidal inferred magnetic fields. The model of a kpc-scale weakly collimated jet/lobe in Mrk\,231 is strengthened by its C-shaped morphology, steep spectral index throughout, complexities in the magnetic field structures, and the presence of self-similar structures observed on the 10-parsec-scale in the literature. The ``wind'' may comprise both nuclear starburst (close to the core) and AGN winds, where the latter maybe the primary contributor. Moving away from the core, the ``wind'' component may also comprise the outer layers (or ``sheath'') of a broadened jet. The inferred value of the (weakly collimated) jet production efficiency, $\eta_\mathrm{jet}\sim$0.01 is consistent with the estimates in the literature. The composite jet and wind outflow in Mrk\,231 appears to be low-power and matter-dominated, and oriented at a small angle to our line of sight.
\end{abstract}




\section{Introduction}
\label{sec:intro}
Active galactic nuclei (AGN) are powered by the release of gravitational energy as matter is accreted onto supermassive ($10^6-10^{10}$~M$_{\sun}$) black holes. Seyfert galaxies are a sub-class of AGN, mostly hosted by spiral or lenticular galaxies. They are typically classified as ``radio-quiet'' AGN as their ``radio-loudness'' parameter $R\equiv \mathrm{S_{radio}/S_{optical}}$ is often less than 10 \citep{Kellermann89}; S is the flux density at radio (5~GHz) and optical (B-band) frequencies. Many Seyfert galaxies however do not adhere to the formal dividing line and move into the ``radio-loud'' category when the galactic optical contribution is properly accounted for \citep[e.g.,][]{HoPeng01,Kharb12a}. Seyferts have been further classified as type 1 or type 2 depending on the presence of broad and narrow emission lines versus only narrow emission lines \citep{Antonucci93}. 

The origin of radio outflows in Seyfert galaxies is still a matter of debate, with AGN jets/winds and starburst superwinds being the primary contenders \citep{Ulvestad81,Baum93,Colbert96,Panessa19}. \citet{Sebastian19a,Sebastian19b,Sebastian20} have used radio polarimetry to compare the radio outflows of Seyfert galaxies to starburst galaxies and found suggestions of differences in their degree of polarization. \citet{Silpa21} have deduced that the Seyfert 1 galaxy III~Zw~2 harbours a composite jet and wind outflow exhibiting toroidal inferred magnetic (B-) fields, from their radio polarimetric study.

Mrk\,231 ({\it aka} UGC~8058 or VII~Zw~490) at a redshift of 0.04217 is a well-studied Seyfert type 1 galaxy \citep{Sanders88b, Surace98}. In fact this optically identified source is known to be the most luminous object in the local ($z<0.1$) Universe \citep{Sanders88b}. It has alternately been classified as a Broad Absorption Line Quasi Stellar Object \citep[BALQSO;][]{Boksenberg77, Smith95, Gallagher02},
and an Ultra Luminous InfraRed Galaxy \citep[ULIRG;][]{Sanders88a}. The synchrotron emission from Mrk\,231 is known to be variable on a timescale of a few days \citep{McCutcheonGregory78} to years \citep{Condon91b}. Mrk\,231 contains an OH megamaser \citep{Baan85} as well as a massive starburst \citep{NeffUlvestad88}. Very Long Baseline Array (VLBA) observations of Mrk\,231 have revealed a triple radio source $\sim2$ parsec in extent, with a PA $\sim 5\degr$ \citep{Ulvestad99a}. 

Mrk 231 exhibits AGN-driven outflows in multiple gas phases \citep{Feruglio15}
and over multiple spatial scales \citep[e.g.,][]{Veilleux16}. The presence of a parsec-scale AGN jet and a BAL wind has been documented by \citet{Reynolds09}. The study by \citet{RupkeVeilleux11} reveals the presence of an AGN-driven wind, starburst-driven wind and a radio jet in Mrk\,231. \citet{Veilleux16} suggest that Mrk\,231 is the nearest example of  weak-lined ``wind-dominated'' quasars with high accretion rates. The morphology and kinematics of the neutral atomic and molecular gas outflows as presented by \citet{Feruglio15, RupkeVeilleux13, Rupke17} point to a cold gas ``wide angle outflow'' on $\sim1-20$ kpc scales.
The co-existence of a radio jet and powerful multi-phase outflows in Mrk\,231 \citep{Reynolds09, Reynolds17} make it an ideal candidate to study the origin of outflows in AGN.

We present here radio polarimetric observations of Mrk\,231 at multiple frequencies and resolutions with the Very Large Array (VLA). In this paper, we adopt a cosmology with H$_0$ = 73~km~s$^{-1}$~Mpc$^{-1}$, $\Omega_{m}$ = 0.27, $\Omega_{v}$ = 0.73. The spectral index $\alpha$ is defined such that $S_\nu\propto\nu^\alpha$, S$_\nu$ being the flux density at frequency $\nu$.

\begin{table}
\caption{Observation details}
\label{tab:Table1}
\begin{tabular}{cccc}
\hline
VLA Array & Frequency & Observation &
Resolution \\
Configuration & (GHz) & Period &
(arcsec) \\
\hline
A & 1.4 & 1995 July 3 $-$ July 4  & 1.2 \\ 
C & 1.4 & 1996 Feb 17 & 13.4 \\
D & 1.4 &  1995 May 12 $-$ June 2 & 40.2 \\
A & 4.9 &  1995 July 21 $-$ July 22 & 0.3 \\
B & 4.9 & 1995 Nov 2 & 1.2 \\
C & 4.9 & 1996 Feb 17 & 4.0 \\
D & 4.9 & 1995 May 3 $-$ May 24 & 11.9 \\
\hline
\end{tabular}
\end{table}

\section{Radio Data Analysis}
We observed Mrk\,231 with the VLA at 1.4 and 4.9~GHz in the A, B, C, and D-array configurations during 1995 - 1996 (Project ID: AB740). Table~\ref{tab:Table1} provides the details of the observations. 3C286 and 3C48 were used as the primary flux density as well as the polarisation calibrators, while J1400+621 was used as the phase calibrator for the whole experiment. The data were processed with AIPS using standard imaging and self-calibration procedures. AIPS task PCAL was used to solve for the antenna ``leakage'' terms (D-terms) as well as polarisation of the calibrators 3C286 and 3C48. The leakage amplitudes were typically a few percent. Polarisation calibration of the 4.9~GHz A-array data was unsuccessful.\footnote{This was a consequence of not having acquired enough scans of the polarisation calibrator for adequate parallactic angle coverage; there were only two scans of 3C286. Moreover, the polarization properties of the phase calibrator J1400+621 are not well-known.}

The polarisation intensity {\tt (PPOL)} and polarisation angle {\tt (PANG)} images were created by first running the {\tt AIPS} task {\tt IMAGR} for Stokes `Q' and `U' and then combining these images using the task COMB. Pixels with intensity values below 3$\sigma$ and angle errors $>10\degr$ were blanked before making {\tt PPOL} and {\tt PANG} images, respectively. (The cut on the angle error was relaxed to 15$\degr$ for the 1.4~GHz A-array image.) Fractional polarisation images were created using the {\tt PPOL} and total intensity images where pixels with $>10\%$ errors were blanked. 

We created two two-frequency spectral index images using the {\tt AIPS} task {\tt COMB} after convolving images at both frequencies with the same circular beams (1.5$\arcsec$ and 15.0$\arcsec$). Pixels with total intensity value below 3$\sigma$ were blanked before making the spectral index images. Rotation measure (RM) images were created using three frequencies with {\tt AIPS} tasks {\tt MCUBE}, {\tt TRANS}, and {\tt RM}. For this, we split the two IF data at L-band (IF1 = 1.465~GHz, IF2 = 1.385~GHz) and used the C-band (frequency = 4.860~GHz) data as such. Two RM images were made after convolving all images with the same circular beam (1.5$\arcsec$ and 15.0$\arcsec$). 

Flux density values reported in the paper were obtained using the Gaussian-fitting {\tt AIPS} task {\tt JMFIT} for compact components like the core, and {\tt AIPS} task {\tt TVSTAT} for extended emission. The rms noise values were similarly obtained using {\tt AIPS} tasks {\tt TVWIN} and {\tt IMSTAT}.

\begin{figure}
\centerline{
\includegraphics[height=9.8cm,trim=65 160 0 100]{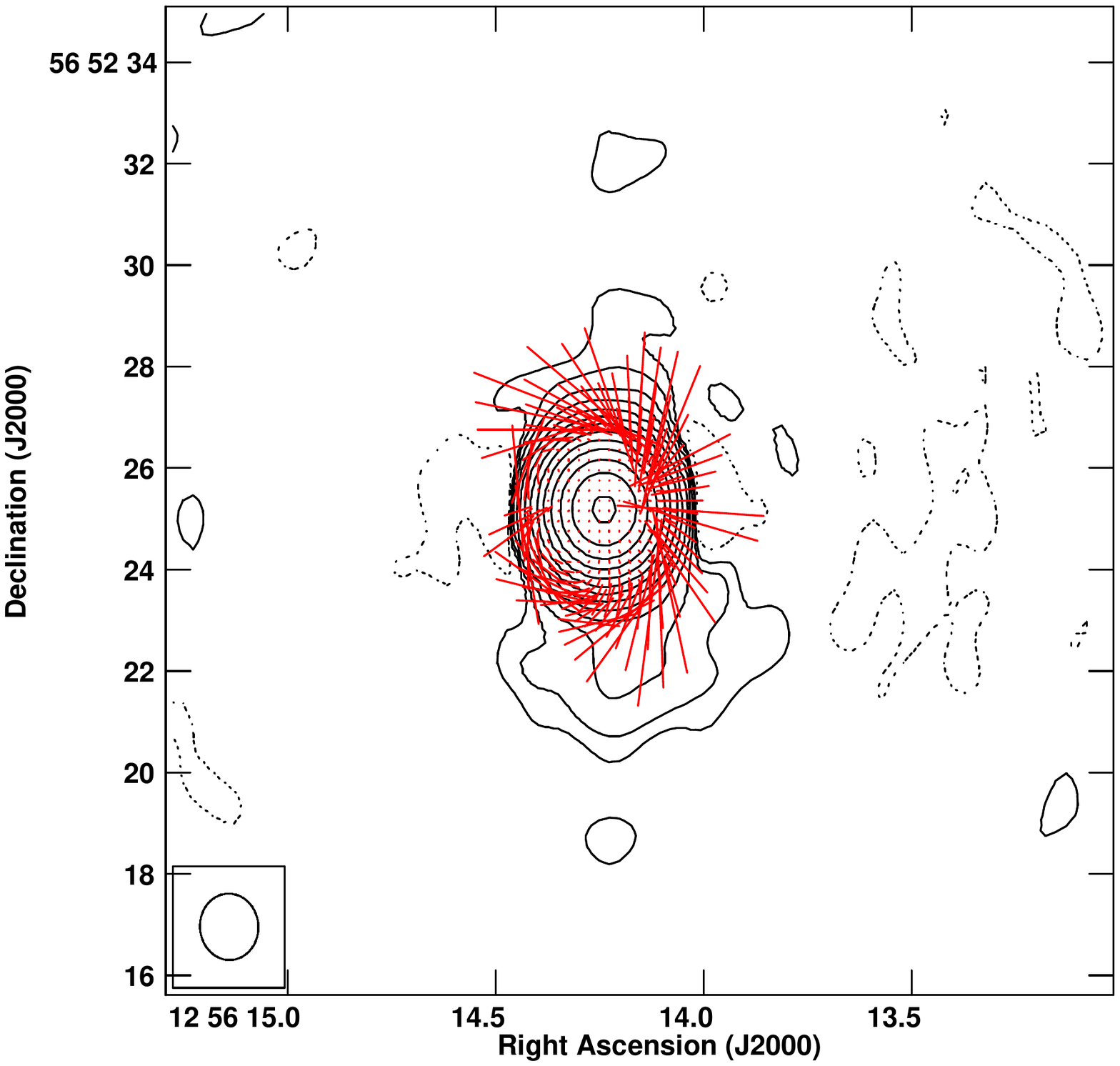}}
\caption{\small 4.86~GHz VLA B-array total intensity contour image with electric polarisation vectors in red. Contour levels are in percentage of peak surface brightness (=250.2 mJy~beam$^{-1}$) and increase in steps of 2, with the lowest contour level being at $\pm0.021$\%. Polarization vector of length 1$\arcsec$ corresponds to fractional polarization of 41.7\%.}
\label{fig1}
\end{figure}

\section{Results}
\subsection{Radio morphology}
A bright core and fainter emission with a north-south orientation, extending to $\sim$4 kpc south from the core, is detected in the 4.9~GHz B-array VLA image of Mrk\,231 (Figure~\ref{fig1}). The 1.4~GHz A-array image reveals a $\sim$ 25 kpc-scale poorly collimated, filamentary radio structure to the south, extending beyond the optical emission of the host galaxy (Figure~\ref{fig2}, left panel). See also the VLA 1.4~GHz A-array image of Mrk\,231 in \citet{Morganti16}. Specifically, the resemblance of Fig. 2 left in the current work to Fig. 6 in \citet{Morganti16} suggests that the lobe morphology is real, and not completely affected by imaging artefacts, although some of the ``ribbed'' features might indeed by artefacts since we see them on the counter-lobe side. Similar filamentary radio lobes have been observed in other Seyfert galaxies \citep[e.g., NGC 3079;][]{Sebastian19a}. The 4.9~GHz C-array image reveals diffuse lobe-like emission to the south that extends to $\sim$ 30 kpc (Figure~\ref{fig2}, right panel), which was also detected in the 1.4 GHz images of \citet{Ulvestad99a}. The 1.4 GHz C-array and 4.9 GHz D-array images reveal significant amounts of diffuse emission around the core and diffuse lobe-like emission extending $\sim$55~kpc to the south (Figure~\ref{fig3}). The 4.9~GHz D-array image also shows an extension towards the north. The broad radio structures seen towards the south in Figure~\ref{fig3} encompasses the 
radio structures revealed in Figure~\ref{fig2}. 

The left panel of Figure~\ref{fig4} shows that the region where the outflow in the 1.4 GHz A-array image de-collimates and spreads out to the south-west, coincides with the edge of the optical diffuse emission. The outflow in the 1.4 GHz C-array image also shows a broad curvature towards the west, which in fact, coincides with the direction of the bending of the tidal arm in the south (Figure~\ref{fig4}, right panel). The morphology and the extent of the radio structure in the 1.4~GHz C-array image resemble the WSRT image presented in \citet{Morganti16}. The 1.4~GHz D-array image in Figure~\ref{fig5} is the lowest resolution image of Mrk\,231 from the current project. The extent of the radio structure in this image is $\sim3\arcmin$ ($\sim145$~kpc). \citet{Reynolds09} have deduced that the outflow inclination is small on the basis of slow Very Long Baseline Array (VLBA) speeds. This is consistent both with its BALQSO nature as well as its kpc-scale morphology as seen in the left panel of Figure~\ref{fig3}. The 4.9 GHz A-array image reveals only a radio core.

Based on our multi-resolution images we find that the kpc-scale radio structure to the south in Mrk\,231 resembles a ``flaring jet'' or lobe, rather than a radio ``bubble''. Flaring jets leading into lobes are typically observed in Fanaroff-Riley type I (FRI) radio galaxies \citep[e.g.,][]{Bondi01,LaingBridle02}. The radio structure is one-sided and C-shaped (see Figure~\ref{fig3}, left), and shows directionality rather than spherical symmetry expected from a bubble. The spectral index images do not show any change in values but is steep spectrum throughout. For instance, a spectral index flattening is observed at the edges of the radio bubbles in the Seyfert galaxy NGC\,6764 \citep{Hota06, Kharb10}. It is interesting to note that the 10-kpc-scale and the 10-parsec-scale radio structures in Mrk\,231 appear to be ``self-similar'' \citep[see Figure~1 in][]{Morganti16}, reminiscent of the lobes observed in Mrk\,6 \citep{Kharb06}. In the case of Mrk\,231, the radio structures on the two scales are not as clearly delineated, most likely due to their orientation being close to our line of sight. The presence of the two distinct structures may even be consistent with episodic jet activity in Mrk\,231, like in Mrk\,6, but cannot be confirmed without additional spectral index data on multiple scales. Thus, based on the observational evidence, we have adopted the interpretation that we see a weakly collimated jet on the kpc scale. With this general scenario in mind, we have estimated the ``jet power'' of Mrk\,231 in Section 3.5.1 ahead.


\subsection{Polarization structures}
\label{sec3.2}
\begin{figure*}
\centerline{
\includegraphics[height=9.5cm,trim=50 140 65 120]{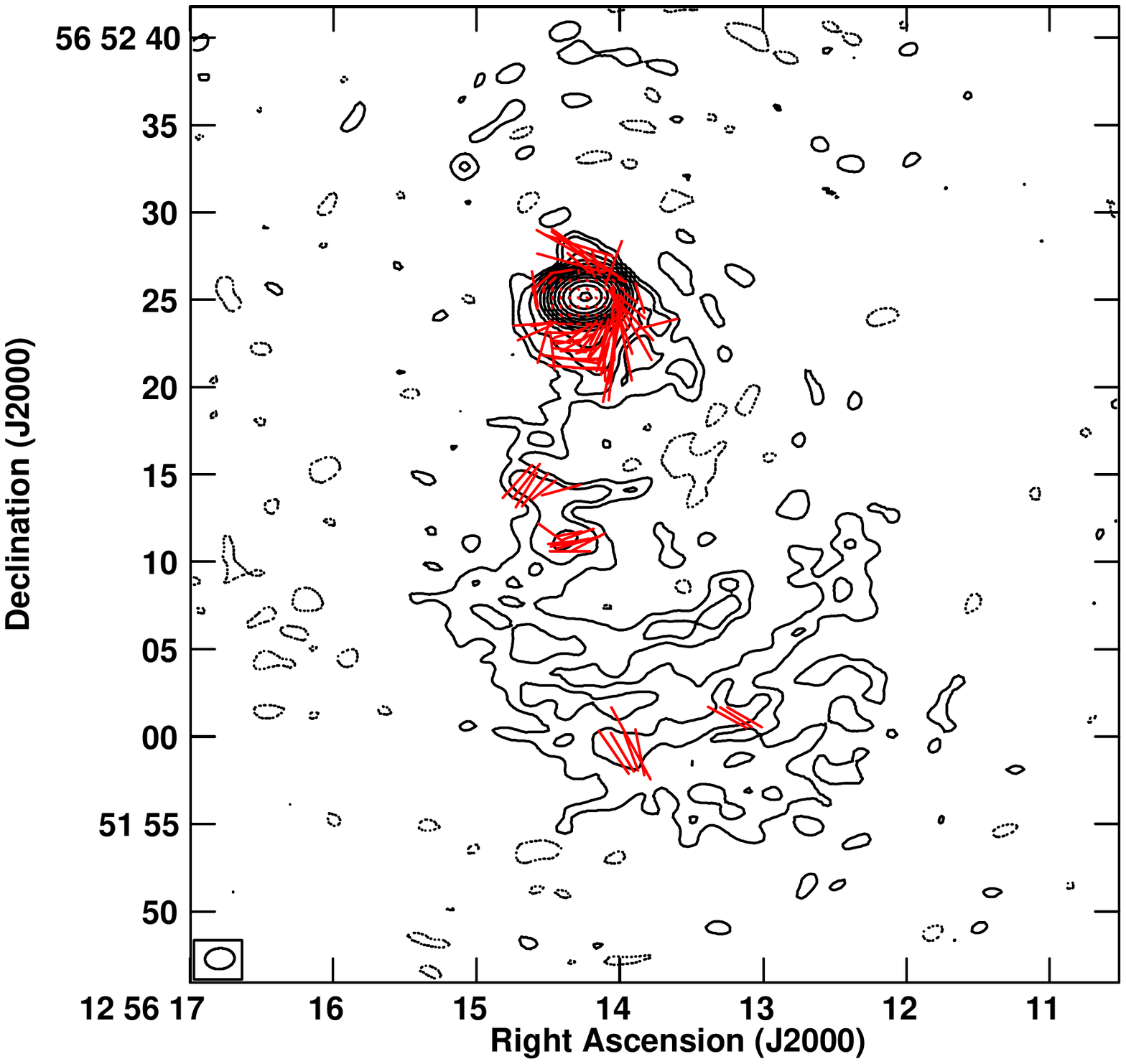}
\includegraphics[height=10cm,trim=20 160 0 120]{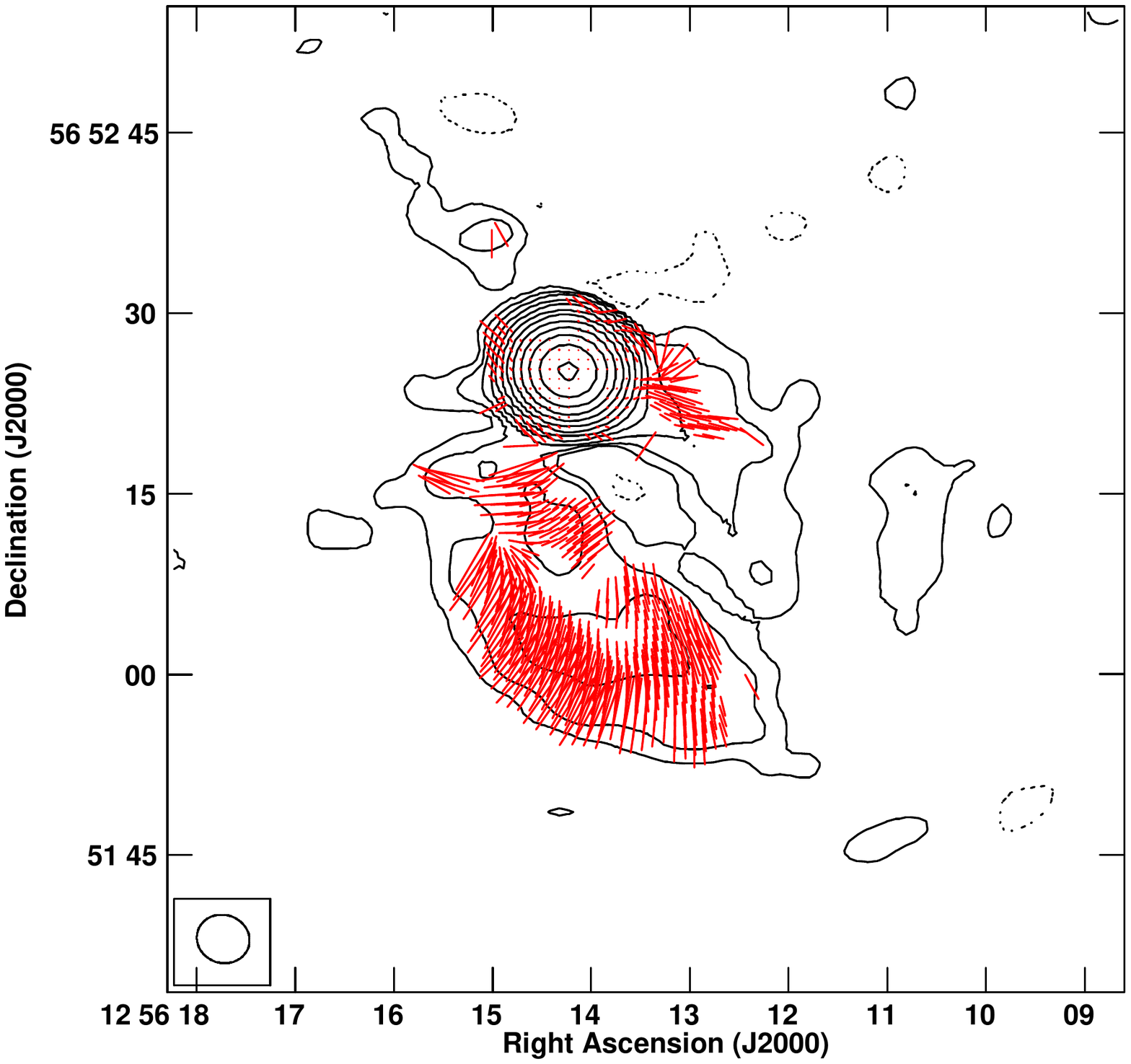}}
\caption{\small Left: 1.42~GHz VLA A-array total intensity contour image with electric polarisation vectors in red. Contour levels are in percentage of peak surface brightness (=230.95~mJy~beam$^{-1}$), with contours levels increasing in steps of 2, with the lowest contour level being $\pm0.020$\%. Polarization vector of length 1$\arcsec$ corresponds to fractional polarization of 20\%. Right: 4.86~GHz VLA C-array total intensity contour image with polarisation vectors in red. Contour levels are in percentage of peak surface brightness (=258.3~mJy~beam$^{-1}$), with contours levels increasing in steps of 2, with the lowest contour level being $\pm0.042$\%. Polarization vector of length 1$\arcsec$ corresponds to fractional polarization of 12.5\%.}
\label{fig2}
\end{figure*}

\begin{figure*}
\centerline{
\includegraphics[height=8cm,trim=150 200 0 160]{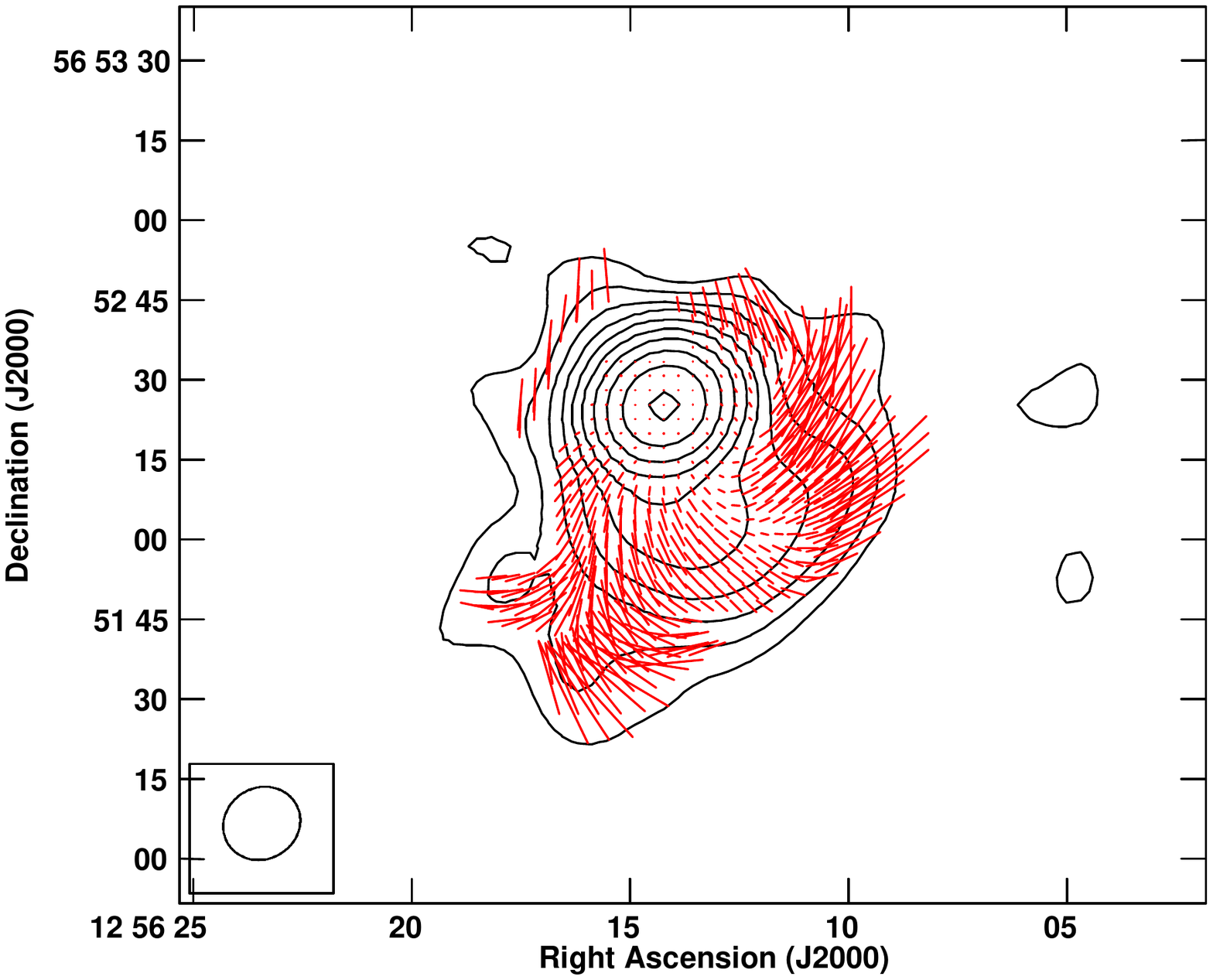}
\includegraphics[height=8.5cm,trim=70 235 100 160]{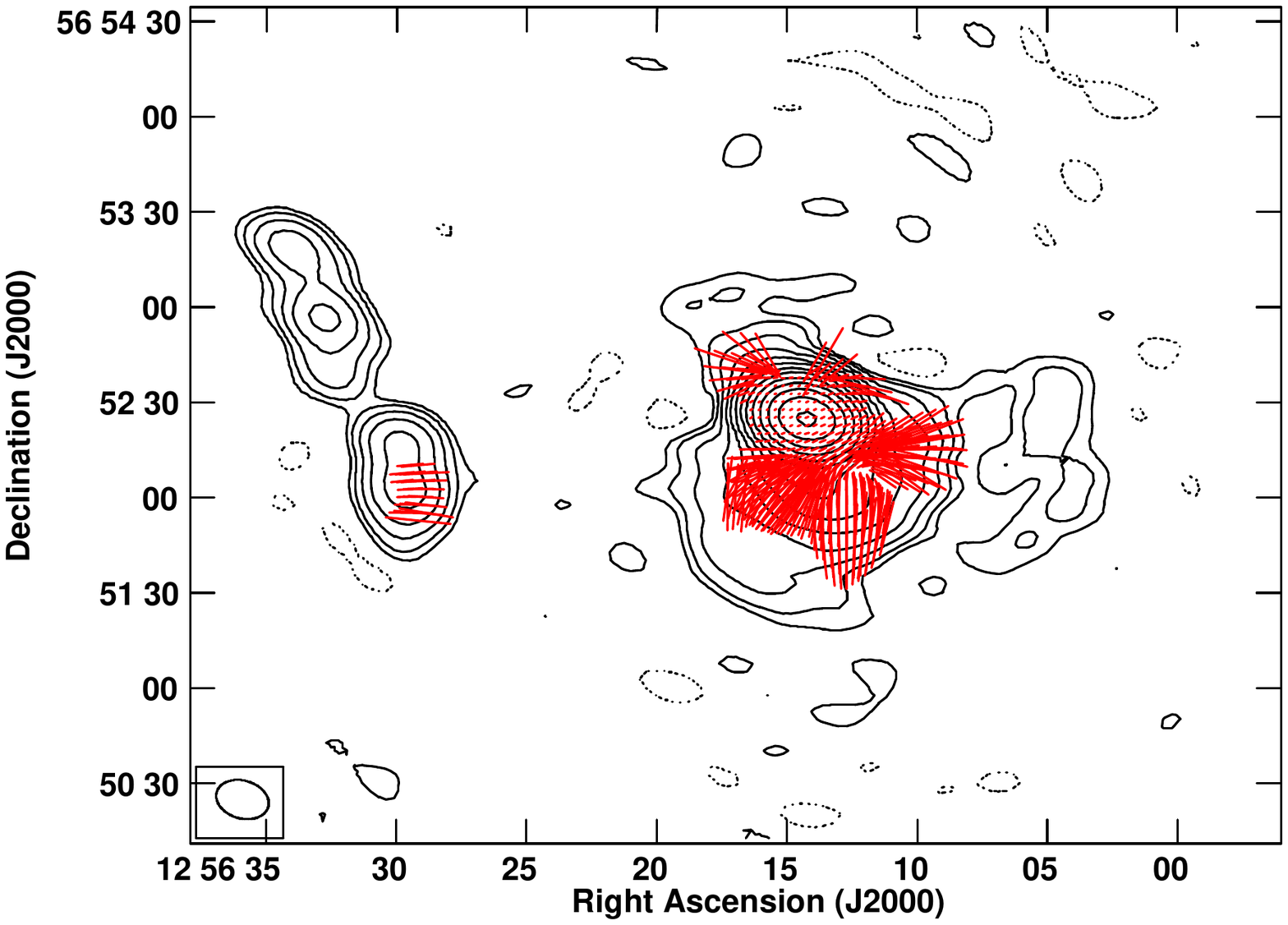}}
\caption{\small Left: 1.42~GHz VLA C-array total intensity contour image with electric polarisation vectors in red. Contour levels are in percentage of peak surface brightness (=252.1~mJy~beam$^{-1}$), with contours levels increasing in steps of 2, with the lowest contour level being $\pm0.17$\%. Polarization vector of length 1$\arcsec$ corresponds to fractional polarization of 3.7\%. Right: 4.86 GHz VLA D-array total intensity contour image with polarisation vectors in red. Contour levels are in percentage of peak surface brightness (=270.2~mJy~beam$^{-1}$), with contours levels increasing in steps of 2, with the lowest contour level being $\pm0.021$\%. Polarization vector of length 1$\arcsec$ corresponds to fractional polarization of 1.3\%.}
\label{fig3}
\end{figure*}

\begin{figure*}
\centering{
\includegraphics[width=9cm,trim=0 120 0 160]{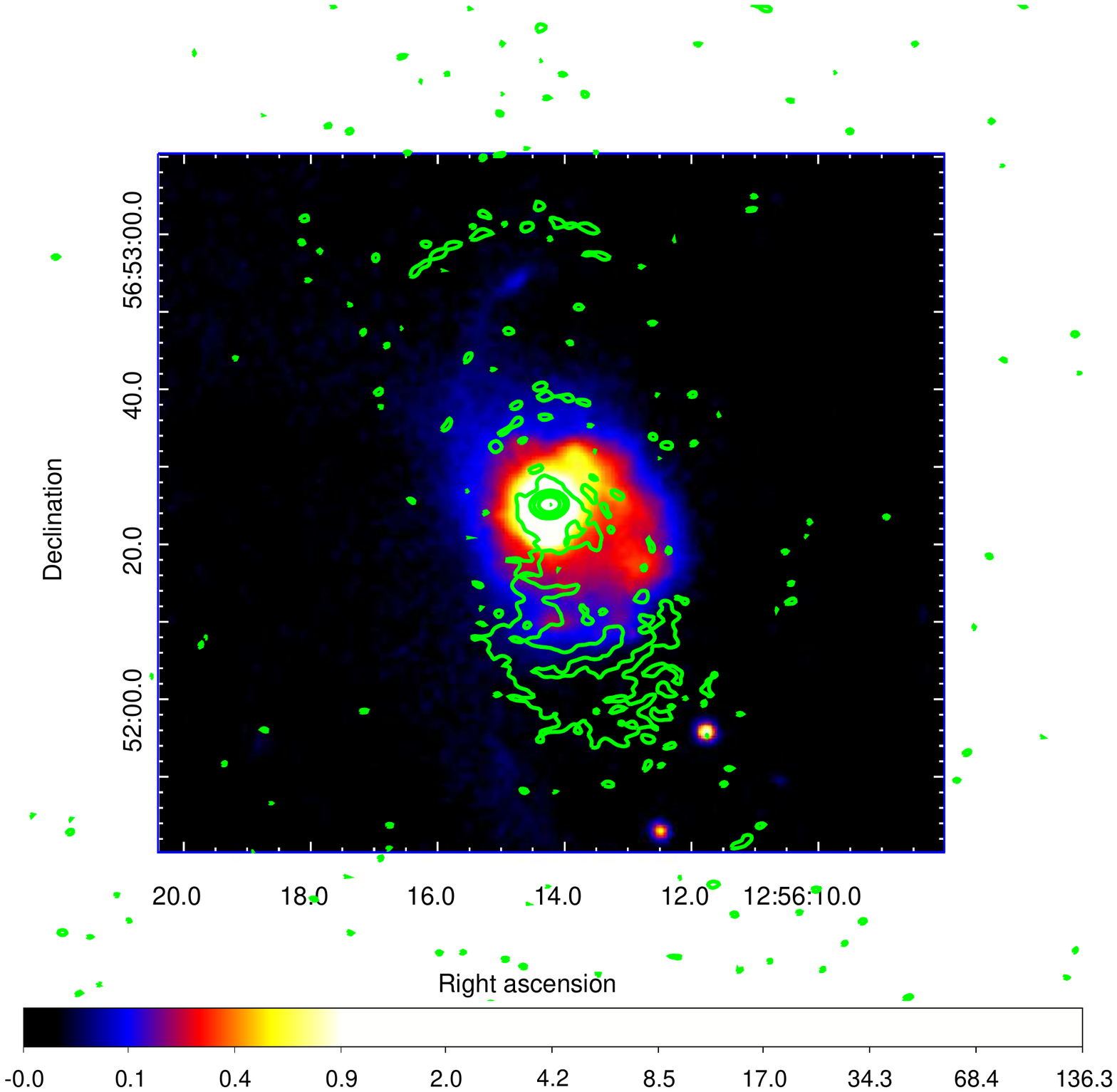}
\includegraphics[width=9cm,trim=0 120 0 160]{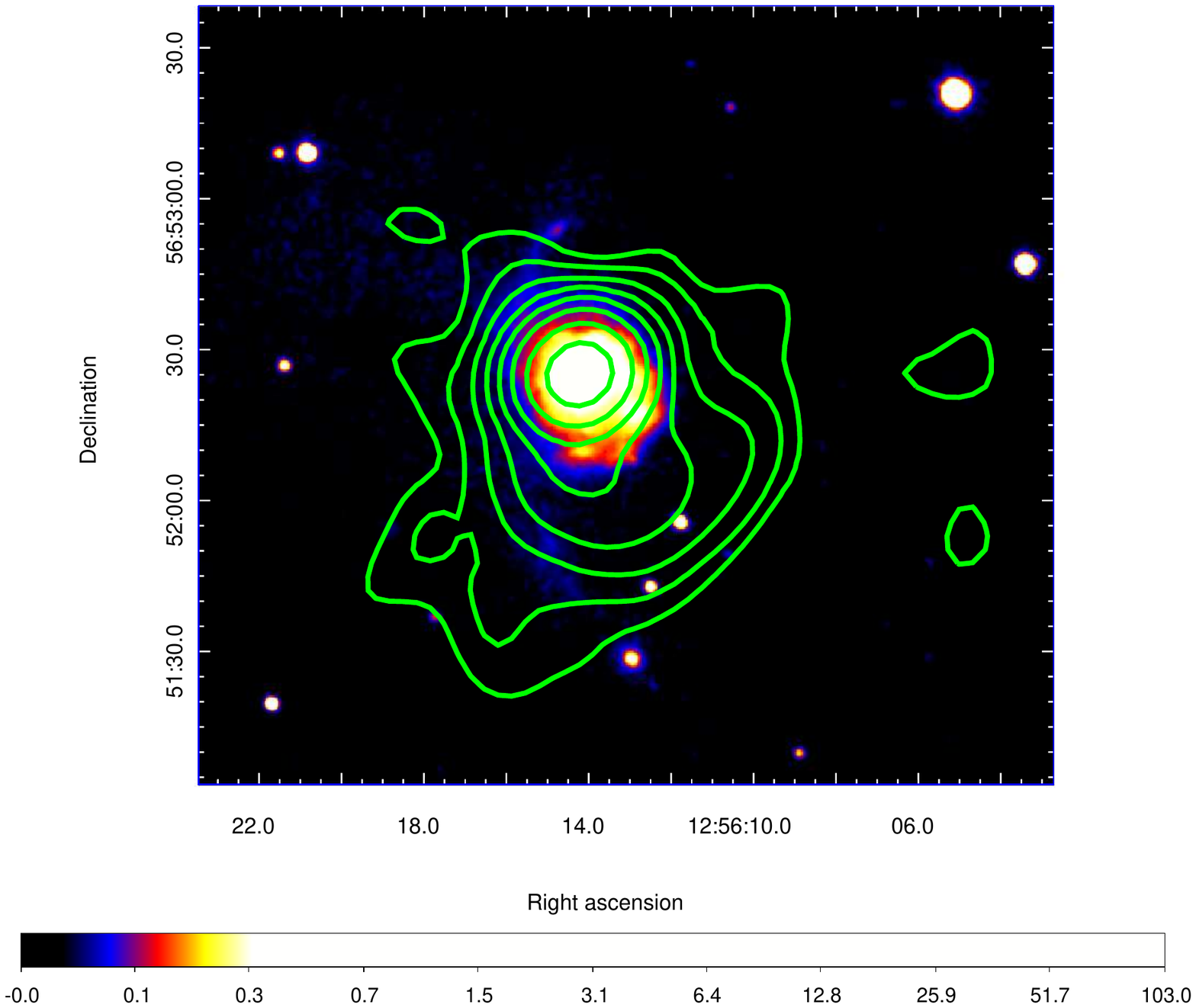}}
\caption{\small Left: 1.42 GHz VLA A-array total intensity contours overlaid on SDSS \citep{Ahumada20} $\it{g}$-band optical image in color. The color scale extends from 0 to 136.3 $\times$ 14~nJy or 0~nJy to 1.9~$\mu$Jy. Right: 1.42 GHz C-array total intensity contours overlaid on SDSS $\it{g}$-band optical image in color. The color scale extends from 0 to 103 $\times$ 14~nJy or 0~nJy to 1.4~$\mu$Jy.
}
\label{fig4}
\end{figure*}

Complex polarization structures are revealed in the VLA images. Figures~\ref{fig1} $-$ \ref{fig3} and \ref{fig5}, which are presented in the decreasing order of resolution, show the polarization electric ($\chi$) vectors as red ticks with lengths proportional to linear fractional polarization. The rotating $\chi$ vectors (with errors of a few degrees) at the core edges in Figure~\ref{fig1} suggest tangential or compressed B-fields, assuming optically thin emission.

The kpc-scale weakly collimated jet in the south shows low levels of polarization in the higher resolution 1.4 GHz A-array image (Figure~\ref{fig2}, left panel) whereas strong polarization in the lower resolution 4.9 GHz C-array image (Figure~\ref{fig2}, right panel). This can be explained as an effect of wavelength-dependent depolarization at 1.4 GHz \citep{Burn66}. We find four distinct polarized regions along the weakly collimated jet in the 1.4 GHz A-array image.
The regions closer to the core exhibit inferred B-fields parallel to the local jet direction whereas the regions further away exhibit inferred B-fields aligned with the lobe edges, assuming optically thin emission. 
The 4.9 GHz C-array image also reveals inferred B-fields aligned with the outer-edges of the lobe and a region closer to the core with inferred B-fields parallel to the local jet direction.
It also reveals a region to the west of the core with inferred transverse B-fields. The lower polarization of the core in the 4.9~GHz C-array image as compared to the 1.4~GHz A-array image can be explained as a beam depolarization effect.

The B-field structure in the 1.4 GHz C-array image is not representative of a typical ``bubble'' with aligned fields at the edges, but rather shows complexities. The 4.9 GHz D-array image reveals three distinct regions of ordered B-fields in the southern side and two distinct regions of ordered B-fields in the northern side of the core (Figure~\ref{fig3}, right panel). The transverse inferred B-fields in the core of the 4.9~GHz B-array image continues all the way up to the edge of the lobe as seen in the 1.4~GHz C-array and 4.9~GHz D-array images. The B-fields in the 1.4~GHz C-array image also clearly follow the curvature (C-shape) in the lobe, similar to that observed in the south-western lobe of III~Zw~2 \citep{Silpa21}. The 1.4 GHz D-array image reveals ordered B-fields at the centre and aligned fields at the edges (Figure~\ref{fig5}). We also note that had the kpc-scale radio structure been a typical radio ``bubble'', different resolution images at the same frequency would have yielded similar polarization structures except for differences arising from beam depolarization effects, which is not the case of Mrk\,231.

The total flux density, polarized flux density, fractional polarization (F.P) and polarization angle values for different radio components of Mrk\,231 at different resolutions are summarized in Table~\ref{tab:Table2}. The F.P values in the outflow (jet and jet+wind) are typically $40-50$\% in the higher resolution images and $20-30$\% in the lower resolution images at both frequencies. The F.P value for the composite core+jet+wind radio structure in the D-array 1.4 GHz image is $\sim17\%$.

\subsection{A rotation measure analysis}
\label{sec3.3}
The $1.4-4.9$~GHz rotation measure (RM) images of Mrk\,231 at resolutions of 15$\arcsec$ and 1.5$\arcsec$ are presented in the left and right panels of Figure~\ref{fig6}, respectively. The mean RM values are $14\pm2$ and $20\pm4$ rad~m$^{-2}$ (errors are obtained using error propagation rules) in the 15 arcsec
image and 1.5 arcsec image, respectively. The three inlayed plots in the left panel of Figure~\ref{fig6} present the radial RM profiles along the slices AB, BC and CD. The RM is given by the following relation:
\begin{equation}
\mathrm{RM = \int 812\,n_e\,B_{||}\,dl~~~rad~m^{-2}},
\end{equation}
where n$_\mathrm{e}$ is the electron number density of the Faraday rotating medium in cm$^{-3}$, B$_{||}$ is the line-of-sight component of the B-field in milliGauss (mG) and dl is the path length of the Faraday screen in parsecs (pc). Based on the above relation, the radial RM profile can translate to a radial electron number density (n$_\mathrm{e}$) profile of the Faraday rotating medium through which the jet passes, assuming a constant B$_{||}$ field and that the Faraday rotating medium has the same B-field as the radio source. We also assume that internal Faraday rotation is not a significant contributor. The B$_{||}$ field could be taken as the equipartition field of $\sim3\times10^{-3}$ mG\footnote{Assuming a cylindrical geometry with volume filling factor and proton-to-electron energy ratio equal to 1.0, a spectral index value of $-0.8$ and the ``minimum'' energy relations from \citet{O'DeaOwen87}.} (using the weakly collimated jet/lobe in 4.9 GHz C-array image).

Using the {\tt PYTHON} fitting procedure {\tt CURVE$\_$FIT}, we obtain the following best fitting power-laws for the radial RM profiles, or equivalently the radial n$_\mathrm{e}$ profiles: n$_\mathrm{e}$ $\propto$ RM $\propto$ r$^\mathrm{a}$, where a = $-0.11$, $-0.17$ and 0.49 for slices AB, BC and CD respectively. Here, r is the radial distance from the core for slice AB and from the point B for slices BC and CD. The three slices were chosen on the basis of a drastic change in the RM profile, as well as the direction of the weakly collimated jet. The pressure profile of the external medium could be inferred from the n$_\mathrm{e}$ profile as p$_\mathrm{gas}$ $\propto$ n$_\mathrm{e}^{\gamma}$, assuming adiabatic equation of state for non-relativistic mono-atomic gas and $\gamma$ = 5/3 being the ratio of specific heats \citep{Park19}. 
Therefore, we infer p$_\mathrm{gas}$ $\propto$ r$^\mathrm{b}$, where b = $-0.18$, $-0.28$ and 0.82 for slices AB, BC and CD respectively.

We know that the magnetic pressure, p$_\mathrm{mag}$ is B$^2$/8$\pi$. The poloidal B-field varies as r$^{-2}$ and the toroidal B-field varies as r$^{-1}$ v$^{-1}$ \citep{Begelman84}. The magnetization parameter, $\beta$, is given as the ratio of gas pressure to magnetic pressure, i.e.
\begin{equation}
\mathrm{\beta = p_{gas}/p_{mag}}. 
\end{equation}
For poloidal B-fields, we find that $\beta$ $\propto$ r$^\mathrm{c}$, where c = 3.8, 3.7 and 4.8 for slices AB, BC and CD respectively. For toroidal B-fields, we find that $\beta$ $\propto$ r$^\mathrm{d}$, where d = 1.8, 1.7 and 2.8 for slices AB, BC and CD respectively. Thus, for both poloidal and toroidal B-fields, the value of $\beta$ increases with distance from the core. We use this information to infer that the outflow is matter-dominated as discussed ahead in Section~\ref{sec4.4}. One caveat in this analysis is the assumption of self-similarity, which may be too simplistic for this complex system.

\subsection{The extended emission around the radio core}
\label{sec4.1}
The diffuse radio emission extending $\sim$4 kpc from the core in the 4.9 GHz B-array VLA image spatially coincides with an optical arc in the HST images, suggesting stellar origin of this extended emission \citep{RupkeVeilleux11,
RupkeVeilleux13, Morganti16, Rupke17}. \citet{Morganti16} refer to this structure as the ``plateau'' in their VLA 1.4~GHz A-array image. The radio-IR correlation is quantified by a q$_\mathrm{IR}$ parameter \citep{Bell03} which is defined as:
\begin{equation}
\mathrm{q_{IR} = log(L_{IR}/3.75\times10^{12} W) - log(L_{1400}/W~Hz^{-1})},
\end{equation}
where L$_\mathrm{1400}$ is the radio luminosity at 1.4 GHz and L$_\mathrm{IR}$ is the infrared luminosity. \citet{Condon02} and \citet{Bell03} classify sources with q$_\mathrm{IR}<1.8$ as AGN-dominated sources. The 1.4 GHz radio luminosity of Mrk\,231 derived using the 4.9 GHz B-array total flux density assuming $\alpha$ = $-0.8$ turns out to be $2.8\times10^{24}$~W~Hz$^{-1}$. Mrk\,231 has an infrared luminosity of $10^{12.37} \mathrm{L}\odot = 9\times10^{38}$~W \citep{Morganti16}. This gives q$_\mathrm{IR}$ = 1.92, which marginally agrees with the nuclear starbursts contributing to the extended radio emission close to the core. As we see below, the starburst contribution to the extended radio emission is however small.

\subsection{Deriving radio contributions from jet \& wind}
\label{sec3.4}
We discuss below an approach that uses different pieces of work from the literature along with the available multi-frequency multi-scale radio data of Mrk\,231 in order to determine the relative contributions of the weakly collimated jet, AGN-driven wind and starburst-driven wind in its total radio emission. We note here that our data, or indeed the current models and simulations for AGN outflows \citep{Mukherjee18}, cannot clearly distinguish between contributions to the ``wind'' component from an accretion disk wind or the outer layers of a broadened jet (like a jet sheath; see Section~\ref{sec:4.2.2} for details). Therefore, caution must be exercised in order to not over-interpret the estimates obtained ahead.

Essentially, we try to estimate the relative contributions of weakly collimated jet and ``wind'' in the total radio emission by using low-resolution and high-resolution images of Mrk\,231. We suggest that the whole of the radio emission detected in the southern lobe of the 1.4 GHz A-array and 4.9 GHz C-array images of Mrk\,231 (Figure~\ref{fig2}) is powered by a weakly collimated jet, given that the high resolution images are likely to be detecting emission from the spine of a jet, rather than diffuse emission from a ``wind''/jet sheath. As we see below, the contribution from a starburst-driven wind to the radio emission in the higher resolution images of Mrk\,231 is also small (see also Section~\ref{sec:3.4.2} where we infer the same but for lower resolution images).

The star-formation rate (SFR) for Mrk\,231 is $\sim 140$~M$_{\sun}$~yr$^{-1}$ \citep[$\sim$30\% of the bolometric luminosity;][]{Veilleux09}. Using \citet{Bell03} star-formation law (see equation 12 in the current work), the amount of radio emission contributed by star-formation in Mrk\,231 turns out to be 3.5$\times10^{39}$ erg s$^{-1}$. The amount of radio emission estimated from the 1.4 GHz A-array and 4.9~GHz C-array flux densities (assuming $\alpha = -0.8$) are $1.4\times10^{40}$~erg~s$^{-1}$ and $4.1\times10^{40}$~erg ~s$^{-1}$ respectively, which are one order of magnitude higher than that estimated from star-formation. 

Using the radio flux densities from the 1.4~GHz A-array and 4.9~GHz C-array images, and assuming them to be jet-dominated, we estimate the jet production efficiency below in Section~\ref{sec:3.4.1}. After this we discuss in Sections~\ref{sec:3.4.2} and \ref{sec:3.4.3}, the amount of radio emission from starburst-driven and AGN-driven winds.


\subsubsection{Jet production efficiency}
\label{sec:3.4.1}
Mass to energy conversion (E = $\eta_\mathrm{rad}$ M c$^2$ with $\eta_\mathrm{rad}$ being the accretion efficiency) is operational in supermassive black holes as matter accretes onto them and powers the AGN. The luminosity emitted by the central engine would be L = $\dot{\mathrm{E}}$ = $\eta_{\mathrm{rad}}$ $\dot{\mathrm{M}}$ c$^2$, where $\dot{\mathrm{M}}$ is the mass accretion rate and c is the speed of light. The radiative efficiency, $\eta_{\mathrm{rad}}$, which is the fraction of the rest mass energy of the accreted matter that is radiated away is given as \citep{Peterson97}:
\begin{equation}
\mathrm{\eta_{rad} = L_{bol}/}\dot{\mathrm{M}}{\mathrm{c^2}}
\end{equation}
The value of $\eta_{\mathrm{rad}}$ is typically 0.18 for standard thin accretion disk \citep{NovikovThorne73}. Analogous to the radiative power, the kinetic power of the ejecting material, such as jets, could be directly linked to the rest-mass energy of the accreting matter \citep{Livio99,Shankar08}. In that case, the kinetic efficiency for the production of jets, i.e. $\eta_\mathrm{jet}$, which is the fraction of the rest mass energy carried away by the jets, is given as:
\begin{equation}
\mathrm{\eta_{jet} = P_{jet}/}\dot{\mathrm{M}} \mathrm{c^2}
\end{equation}
We discuss below two approaches, one following \citet{MerloniHeinz07} and other following \citet{Willot99}, to determine the kinetic power of the weakly collimated jet, P$_\mathrm{jet}$ in Mrk\,231.

(i) We use the empirical relation obtained by \citet{MerloniHeinz07} between P$_\mathrm{jet}$ and 5 GHz radio core luminosity L$_\mathrm{5~GHz}$ for a sample of low-luminosity radio galaxies to derive P$_\mathrm{jet}$:
\begin{equation}
\mathrm{logP_{jet} = 0.81~log(L_{5~GHz}) + 11.9}, 
\end{equation}
The radio core flux density S$_\mathrm{5~GHz}$ of 261~mJy is estimated from the 4.9~GHz~C-array image. L$_\mathrm{5~GHz}$ (in erg~s$^{-1}$) is estimated using:
\begin{equation}
\mathrm{L_{5~GHz} = S_{5~GHz}\times4\pi D_L^2\times10^{7}\times5\times10^{9}}. 
\end{equation} 
Using this value for L$_\mathrm{5~GHz}$ in equation (6), we estimate P$_\mathrm{jet}$ as 7.5$\times$10$^{44}$~erg~s$^{-1}$. This yields $\eta_\mathrm{jet}$ = 0.012, using equation (5)

(ii) Alternately, P$_\mathrm{jet}$ could be estimated using the relation from \citet{Willot99}: 
\begin{equation}
\mathrm{P_{jet} = 3\times10^{38}~L_{151}^{6/7}~f^{3/2}~W}, 
\end{equation}

where L$_{151}$ is the 151~MHz radio luminosity 
in units of $10^{28}$~W~Hz$^{-1}$~sr$^{-1}$ and f, which lies in the range $\sim1-20$, accounts for errors in the model assumptions. \citet{Willot99} make use of the tight correlation between jet kinetic power and narrow-line region (NLR) luminosity for radio-powerful sources \citep[see][]{Rawlings91,BaumHeckman89}. This relation has been slightly modified by \citet{Punsly05}, assuming f = 15, to:
\begin{equation}
\mathrm{P_{jet} \approx 1.1\times10^{45}~(S_{151}~Z^2~(1+z)^{1+\alpha})^{6/7}~(f/15)^{3/2}~erg~s^{-1}},
\end{equation}
where Z$\approx$(3.31$-$3.65)[(X$^4-$0.203X$^3$+0.749X$^2$+0.444X+0.205)$^{-0.125}$], X = 1 + z (z being the redshift), and S$_{151}$ is 151 MHz flux density from the lobe in Jy. Assuming $\alpha=-0.8$, we estimate S$_{151}$ from 4.9 and 1.4 GHz flux densities using:
\begin{equation}
\mathrm{S_{151}=S_{\nu}(151/\nu)^{\alpha}},
\end{equation}
where $\nu=4.9$ and 1.4 GHz.

Using 4.9~GHz flux density, we obtain P$_\mathrm{jet}$ = 3.8$\times$10$^{43}$~erg~s$^{-1}$. This yields $\eta_\mathrm{jet} = 6\times10^{-4}$, using equation (5). Similarly, using 1.4 GHz flux density, we obtain P$_\mathrm{jet}$ = 1.7$\times$10$^{43}$~erg~s$^{-1}$. This yields $\eta_\mathrm{jet} = 3\times10^{-4}$, using equation (5). 

\subsubsection{Radio contribution from starburst-driven wind}
\label{sec:3.4.2}
Next we discuss two approaches, one following \citet{Leitherer99} and other following \citet{DallaVecchiaSchaye08}  (for two different initial mass functions, i.e. IMFs), to estimate the kinetic power of a starburst-driven wind, P$_\mathrm{SBwind}$, and subsequently the radio contribution from a starburst-driven wind, S$_\mathrm{SBwind}$ in Mrk\,231.

A typical supernova ejects 10$^{51}$~erg of kinetic energy \citep{Leitherer99, DallaVecchiaSchaye08}. 

(i) According to \citet{Leitherer99}, the kinetic energy in starburst galaxies is supplied by stellar winds and supernovae. The simulations of \citet{Thornton98} suggest that about 10\% of the available kinetic energy is transferred to the ISM while the rest is radiated away. Therefore, the kinetic power of a starburst-driven wind in a galaxy with SFR of $\Psi$ M$_{\sun}$~yr$^{-1}$ is \citep{ZakamskaGreene14}:
\begin{equation}
\mathrm{P_{SBwind} = 7\times10^{41}\times \Psi~erg~s^{-1}}
\end{equation}
The SFR for Mrk\,231 is $\sim 140~ \mathrm{M_{\sun}~yr^{-1}}$ \citep{Veilleux09}. Therefore, P$_\mathrm{SBwind} = 9.8\times10^{43}$~erg~s$^{-1}$. The amount of radio emission produced by the same galaxy is given by the \citet{Bell03} star-formation law:
\begin{equation}
\mathrm{\nu L_\nu[1.4~GHz] = 2.5\times10^{37}\times \Psi~erg~s^{-1}}
\end{equation}
This implies that the efficiency of conversion from kinetic power to radio luminosity for starburst-driven wind is $3.6\times10^{-5}$.
Therefore, the radio luminosity of the starburst-driven wind is:
\begin{equation}
\mathrm{L_{SBwind} = 3.6\times 10^{-5}\times P_{SBwind}}
\end{equation}
This gives L$_\mathrm{SBwind}$ = 3.5$\times$10$^{39}$ erg s$^{-1}$ for Mrk\,231. Substituting this value of L$_\mathrm{SBwind}$ in equation below:
\begin{equation}
\mathrm{S_{SBwind} = L_{SBwind}/(4\pi D_L^2\times10^{7}\times1.4\times10^{9})}
\end{equation}
we estimate S$_\mathrm{SBwind}$ at 1.4~GHz as $\sim$ 64 mJy. Using $S_\nu\propto\nu^\alpha$ and assuming $\alpha=-0.8$, we estimate S$_\mathrm{SBwind}$ at 4.9~GHz as $\sim$ 23 mJy.
\\
\\
(ii) \citet{DallaVecchiaSchaye08} suggest that about 40\% of the initial kinetic energy is carried away by winds and the remaining is assumed to be radiated away. This has also been suggested by other observations and simulations \citep{Sharma14,Veilleux05,StricklandHeckman09}. A core-collapse supernova releases a kinetic energy of $1.8\times10^{49}$~erg~$M_{\sun}^{-1}$ assuming \citet{Chabrier03} IMF and $1.1\times10^{49}$~erg~$M_{\sun}^{-1}$ assuming \citet{Salpeter55} IMF, into the ISM \citep{DallaVecchiaSchaye08}. Therefore, the kinetic power deposited into the ISM via stellar winds for a galaxy with SFR of $\Phi$ M$_{\sun}$~s$^{-1}$ turns out to be:

(a) for \citet{Chabrier03} IMF,
\begin{align}
\mathrm{P_{SBwind}} &= 40\%~\mathrm{of}~1.8\times10^{49}~\mathrm{erg~{M_{\sun}}^{-1}}\\
        &= 0.72\times10^{49}~\mathrm{erg~{M_{\sun}}^{-1}}\\
        &= 0.72\times10^{49}\times\Phi~\mathrm{erg~s^{-1}}
\end{align}
This implies that the efficiency of conversion from kinetic power to radio luminosity for starburst-driven wind is $1.1\times10^{-4}$. Therefore, the radio luminosity of starburst-driven wind is:
\begin{equation}
\mathrm{L_{SBwind} = 1.1\times 10^{-4}\times P_{SBwind}}
\end{equation}
For Mrk\,231, P$_\mathrm{SBwind}$ is 3.2$\times$10$^{43}$ erg s$^{-1}$, L$_\mathrm{SBwind}$ is 3.5$\times$10$^{39}$ erg s$^{-1}$ and S$_\mathrm{SBwind}$ is $\sim$ 64 mJy at 1.4~GHz (using equation 14) and $\sim$ 23 mJy at 4.9~GHz.\\

(b) for \citet{Salpeter55} IMF, 
\begin{align}
\mathrm{P_{SBwind}} &= 40\%~\mathrm{of}~1.1\times10^{49}~\mathrm{erg~M_{\sun}^{-1}}\\
        &= 0.44\times10^{49}~\mathrm{erg~M_{\sun}^{-1}}\\ 
        &= 0.44\times10^{49}\times \Phi~\mathrm{erg~s^{-1}}
\end{align}

This implies that the efficiency of conversion from kinetic power to radio luminosity for starburst-driven wind is 1.8$\times$10$^{-4}$. Therefore, the radio luminosity of starburst-driven wind is:
\begin{equation}
\mathrm{L_{SBwind} = 1.8\times 10^{-4}\times P_{SBwind}}
\end{equation}
For Mrk\,231, P$_\mathrm{SBwind}$ is $2.0 \times 10^{43}$~erg~s$^{-1}$, L$_\mathrm{SBwind}$ is $3.5 \times 10^{39}$~erg~s$^{-1}$ and S$_\mathrm{SBwind}$ is $\sim$ 64 mJy at 1.4~GHz (using equation 14) and $\sim$ 23 mJy at 4.9~GHz. Interestingly, all three methods yield the same value for S$_\mathrm{SBwind}$.

\subsubsection{Radio contribution from AGN-driven wind}
\label{sec:3.4.3}
We assume that 5\% of the AGN bolometric power is injected into the ambient medium as kinetic power that drives outflows of gas. This is in agreement with that suggested by the cosmological models of galaxy evolution in order to reproduce the observed black-hole bulge (mass) relationships \citep{DiMatteo05, Nesvadba17}. Therefore, we assume that the kinetic power of AGN wind is 5\% of the AGN bolometric luminosity. This yields:
\begin{equation}
\mathrm{P_{AGNwind} = 0.05~L_{bol}}, 
\end{equation}
where L$_\mathrm{bol}$ is the AGN bolometric luminosity. From equation (4), this also implies that 
\begin{equation}
\mathrm{P_{AGNwind} = 0.05~\eta_{rad}~}\dot{\mathrm{M}}\mathrm{~c^2}
\end{equation}
For Mrk\,231, L$_\mathrm{bol} = 1.1\times10^{46}$~erg~s$^{-1}$ \citep{Veilleux09}. Therefore, P$_\mathrm{AGNwind} = 5.5\times10^{44}$~erg~s$^{-1}$.

Thermally and radiatively-driven AGN winds drive synchrotron-emitting shocks as they propagate through the ISM of the host galaxies, similar to the starburst-driven winds. Therefore, following \citet{ZakamskaGreene14}, we could assume that the efficiency of converting the kinetic power of the AGN wind into synchrotron emission is similar to that of the starburst-driven wind. Using this assumption, we derive the contribution of radio emission from AGN winds, S$_\mathrm{AGNwind}$ in Mrk\,231. 
\\
\\
(a) The \citet{Leitherer99} model yields the efficiency of conversion from kinetic power to radio luminosity as 3.6$\times$10$^{-5}$. Following equation 13, 
\begin{equation}
\mathrm{L_{AGNwind} = 3.6\times 10^{-5}\times P_{AGNwind}}
\end{equation}
Substituting the value of L$_\mathrm{AGNwind}$ in the equation below, we estimate S$_\mathrm{AGNwind}$:
\begin{equation}
\mathrm{S_{AGNwind} = L_{AGNwind}/(4\pi D_L^2\times10^{7}\times1.4}\times10^{9}), 
\end{equation}
Using P$_\mathrm{AGNwind}$ = 5.5$\times10^{44}$~erg~s$^{-1}$ in equation 25 yields L$_\mathrm{AGNwind}$ = 2.0$\times$10$^{40}$ erg s$^{-1}$ and S$_\mathrm{AGNwind}$ as $\sim$ 360~mJy at 1.4~GHz (using equation 26). Using $S_\nu\propto\nu^\alpha$ and assuming $\alpha=-0.8$, we estimate S$_\mathrm{AGNwind}$ $\sim$130~mJy at 4.9~GHz for Mrk\,231.\\

(b) Using the \citet{DallaVecchiaSchaye08} model,\\

$\bullet$ for \citet{Chabrier03} IMF, the efficiency of conversion from kinetic power to radio luminosity as 1.1$\times$10$^{-4}$. Following equation 18, 
\begin{equation}
\mathrm{L_{AGNwind} = 1.1\times 10^{-4}\times P_{AGNwind}}
\end{equation}
Using P$_\mathrm{AGNwind}$ = 5.5$\times10^{44}$~erg~s$^{-1}$ in equation 27 yields L$_\mathrm{AGNwind}$ = 6.0$\times$10$^{40}$ erg s$^{-1}$ and S$_\mathrm{AGNwind}$ as $\sim$ 1.1~Jy at 1.4~GHz (using equation 26) and $\sim$400~mJy at 4.9~GHz for Mrk\,231.\\

$\bullet$ for \citet{Salpeter55} IMF, the efficiency of conversion from kinetic power to radio luminosity as 1.8$\times$10$^{-4}$. Following equation 22, 
\begin{equation}
\mathrm{L_{AGNwind} = 1.8\times 10^{-4}\times P_{AGNwind}}
\end{equation}
Using P$_\mathrm{AGNwind} = 5.5 \times 10^{44}$~erg~s$^{-1}$ in equation 28 yields L$_\mathrm{AGNwind} = 1.0 \times 10^{41}$~erg~s$^{-1}$ and S$_\mathrm{AGNwind}$ as $\sim$ 1.8~Jy (using equation 26) at 1.4~GHz and $\sim$ 650~mJy at 4.9~GHz for Mrk\,231.

For completeness, we note that if only 0.5\% of the AGN bolometric luminosity is assumed to be available as the kinetic power of AGN winds \citep{HopkinsElvis10}, all the numbers discussed in this Section would be reduced by a factor of 10. For e.g., P$_\mathrm{AGNwind}$ would be $5.5 \times 10^{43}$~erg~s$^{-1}$. At 1.4~GHz, \citet{Leitherer99} model would yield L$_\mathrm{AGNwind} = 2.0 \times 10^{39}$~erg~s$^{-1}$ and S$_\mathrm{AGNwind}$ $\sim36$~mJy. \citet{DallaVecchiaSchaye08} model would yield L$_\mathrm{AGNwind} = 6.0 \times 10^{39}$~erg~s$^{-1}$ and S$_\mathrm{AGNwind} \sim110$~mJy, assuming \citet{Chabrier03} IMF whereas L$_\mathrm{AGNwind} = 1.0 \times 10^{40}$~erg~s$^{-1}$ and S$_\mathrm{AGNwind}$ $\sim180$~mJy, assuming \citet{Salpeter55} IMF. At 4.9~GHz, \citet{Leitherer99} model would yield S$_\mathrm{AGNwind}$ $\sim13$~mJy. \citet{DallaVecchiaSchaye08} model would yield S$_\mathrm{AGNwind} \sim40$~mJy, assuming \citet{Chabrier03} IMF whereas S$_\mathrm{AGNwind}$ $\sim65$~mJy, assuming \citet{Salpeter55} IMF. We discuss the implications of this in Section~\ref{sec:4.2.4} ahead.

\begin{figure}
\centering{
\includegraphics[width=6.2cm,trim=110 180 110 180]{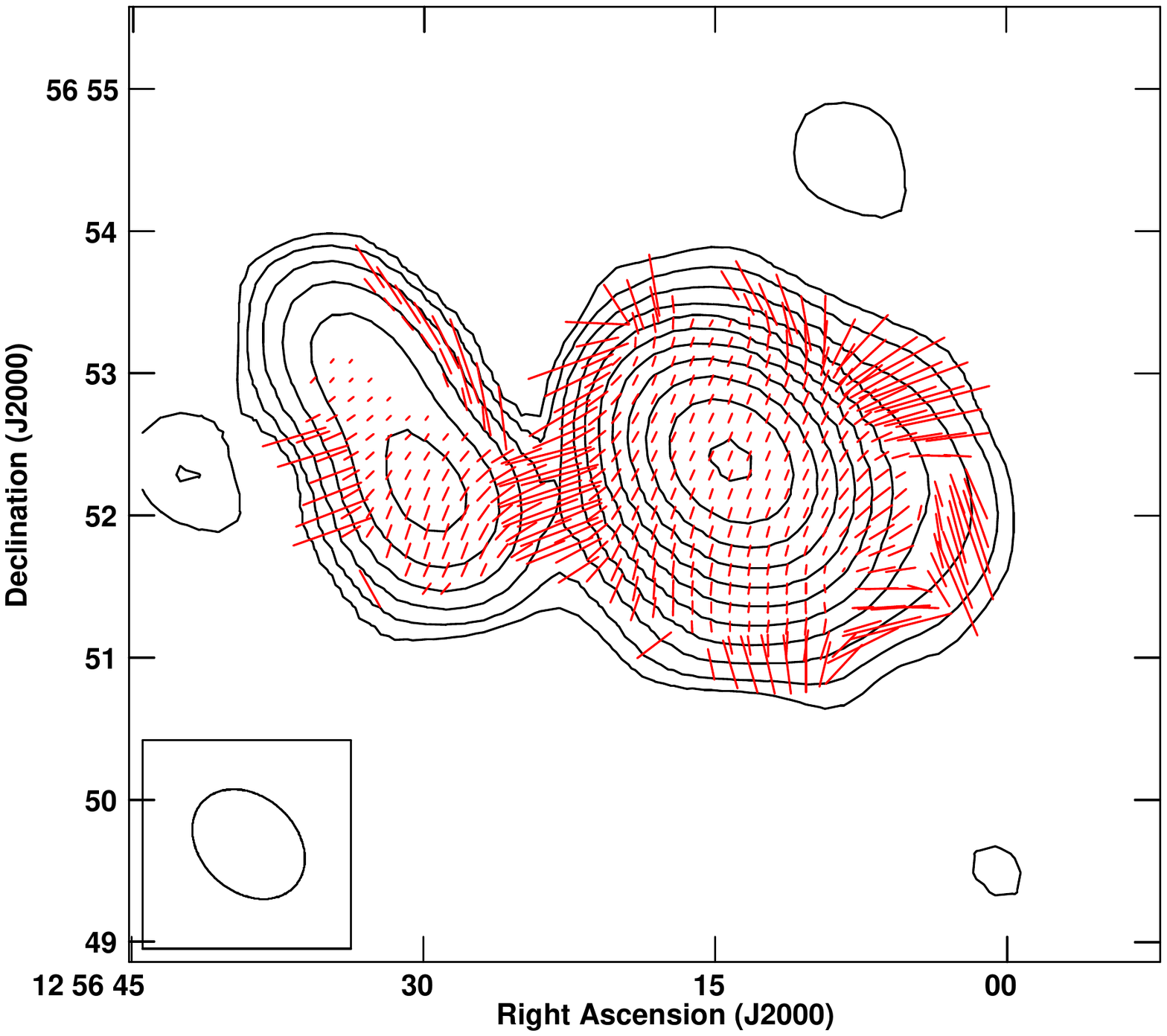}}
\caption{\small 1.42 GHz VLA D-array total intensity contour image with polarisation vectors in red. Contour levels are in percentage of peak surface brightness (=280.3~mJy~beam$^{-1}$), with contours levels increasing in steps of 2, with the lowest contour level being $\pm0.085$\%. Polarization vector of length 1$\arcsec$ corresponds to fractional polarization of 1.6\%.}
\label{fig5}
\end{figure}

\begin{figure*}
\centerline{
\includegraphics[width=8.5cm,trim=100 120 0 180]{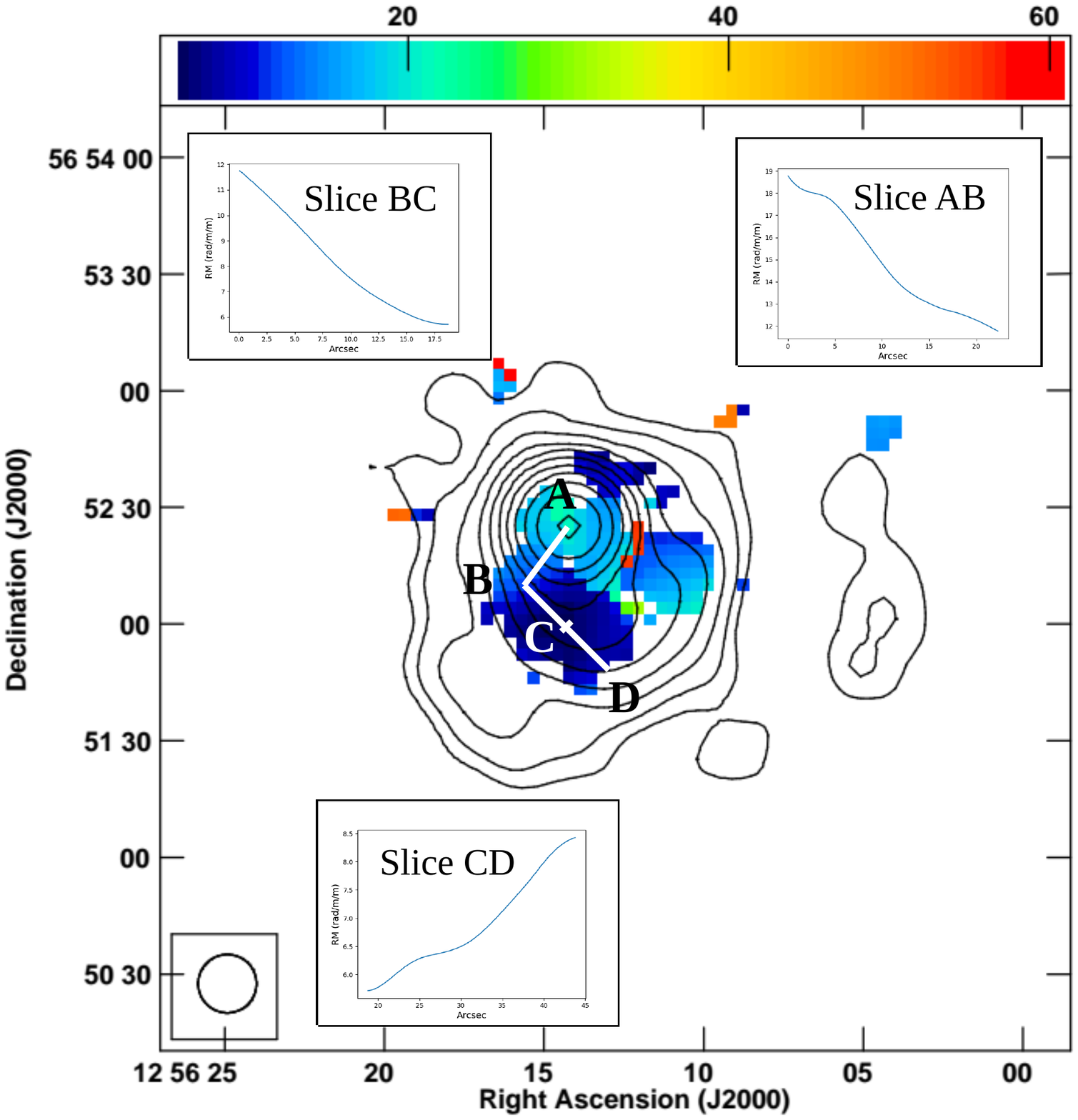}
\includegraphics[width=9.6cm,trim=20 120 10 180]{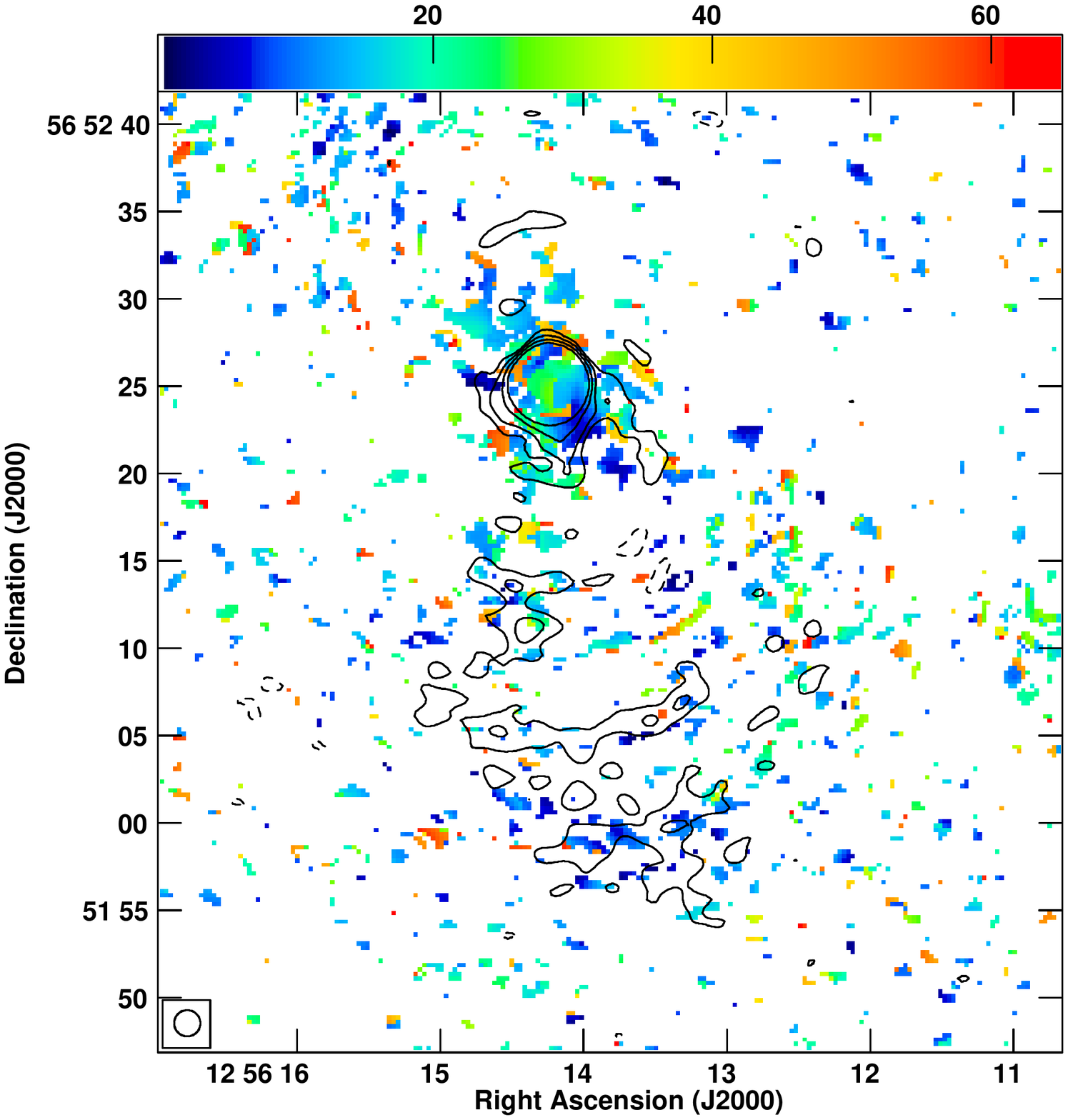}}
\caption{\small The $1.4-4.9$~GHz rotation measure image of Mrk\,231. Left: 15$\arcsec$ resolution image. Contour levels are in percentage of peak surface brightness (=252.7~mJy~beam$^{-1}$), with contours levels increasing in steps of 2, with the lowest contour level being $\pm0.085$\%. The color scale extends from 5.27 to 60.58~rad~m$^{-2}$. Right: $1.5\arcsec$ resolution image. Contour levels are in percentage of peak surface brightness (=232.3~mJy~beam$^{-1}$), with contours levels increasing in steps of 2, with the lowest contour level being $\pm0.042$\%. The color scale extends from 0.55 to 64.78~rad~m$^{-2}$.}
\label{fig6}
\end{figure*}

Assuming that the low-resolution images have contributions from both weakly collimated jet and wind, S$_\mathrm{SBwind}$ and S$_\mathrm{AGNwind}$ would account for the wind contribution from starburst and AGN respectively, and the remaining flux density detected would account for the jet contribution. We use this information to derive the relative contributions of wind and jet in the total radio emission detected in Figures~\ref{fig3} and \ref{fig5}, in Section~\ref{sec:4.2.4}.

\begin{table*}
\begin{center}
\caption{Summary of radio total intensity and polarization properties}
\label{tab:Table2}
\begin{tabular}{lccccc}
\hline
Data & Component & Total flux & Polarized flux & Fractional & Polarization\\
& & density ($I$, mJy) & density ($P$, mJy) & polarization ($P/I$, \%) & angle ($\chi$, $\degr$)\\ \hline
& Core & $246\pm23$ & $4.5\pm0.2$ & $2.1\pm0.1$ & $38\pm1$ \\ 
A-array 1.4 GHz & Jet & 13 & $7\pm1$ & $48\pm10$ & $-7\pm3$ \\ \hline
& Core & $265\pm25$ & $2.5\pm0.3$ & $2.0\pm0.3$ & $-9\pm4$ \\ 
C-array 1.4 GHz & Jet+Wind & 43 & $7.5\pm0.8$ & $29\pm4$ & $-6\pm4$  \\ \hline
D-array 1.4 GHz & Core+Jet+Wind & $318\pm28$ & $22.2\pm0.3$ & $17\pm2$ & $-28\pm3$ \\
\hline
A-array 4.9 GHz & Core & $260\pm24$ & ... & ... & ... \\
\hline
& Core & $260\pm25$ & $5.6\pm0.3$ & $2.8\pm0.2$ & $-3\pm2$ \\ 
B-array 4.9 GHz & Wind & 1.8 & ... & ... & ... \\
\hline
& Core & $261\pm26$ & $1.0\pm0.1$ & $1.2\pm0.2$ & $24\pm6$ \\ 
C-array 4.9 GHz & Jet & 12.0 & $4.0\pm0.6$ & $40\pm7$ & $-19\pm5$ \\ \hline
& Core & $288\pm27$ & $4.6\pm0.1$ & $3.0\pm0.4$ & $-66\pm2$ \\ 
D-array 4.9 GHz & Jet+Wind & 6.2 & $2.7\pm0.2$ & $20\pm3$ & $-22\pm4$ \\
\hline
\end{tabular}
\end{center}
\end{table*}

\section{Discussion}
\label{secdiscussion}
We discuss now the nature of the kpc-scale radio emission in Mrk\,231 and the contributions from various outflow mechanisms.
\subsection{Radio Jet}
\label{sec:4.2.1}
The radio morphology and polarization structures of the 1.4 GHz A-array and 4.9 GHz C-array images (Figure~\ref{fig2}) present a weakly collimated AGN jet exhibiting poloidal inferred B-fields along the spine of the jet \citep[e.g., in III Zw 2;][]{Silpa21}. The radio structure similar to the one revealed in our 1.4 GHz A-array image has also been detected by \citet{Morganti16}; they refer to it as the ``bridge''. Moreover, the total intensity and polarization structure of the 4.9 GHz C-array image resemble the 1.4 GHz images presented in Figures 4a and 5 of \citet{Ulvestad99a}. They suggest that the diffuse lobe emission to the south is non-stellar in origin and is powered by a radio jet. 
They also attribute the wrapping of the B-fields around the outer edges of the lobe to jet emission. 
Similar B-field morphology is observed in the southern lobe of our 4.9 GHz C-array image but with an offset in the polarization angles at some places and consistent at other places, suggestive of a non-uniform Faraday screen. 

\begin{figure*}
\centerline{
\includegraphics[width=7.8cm,trim=130 180 10 160]{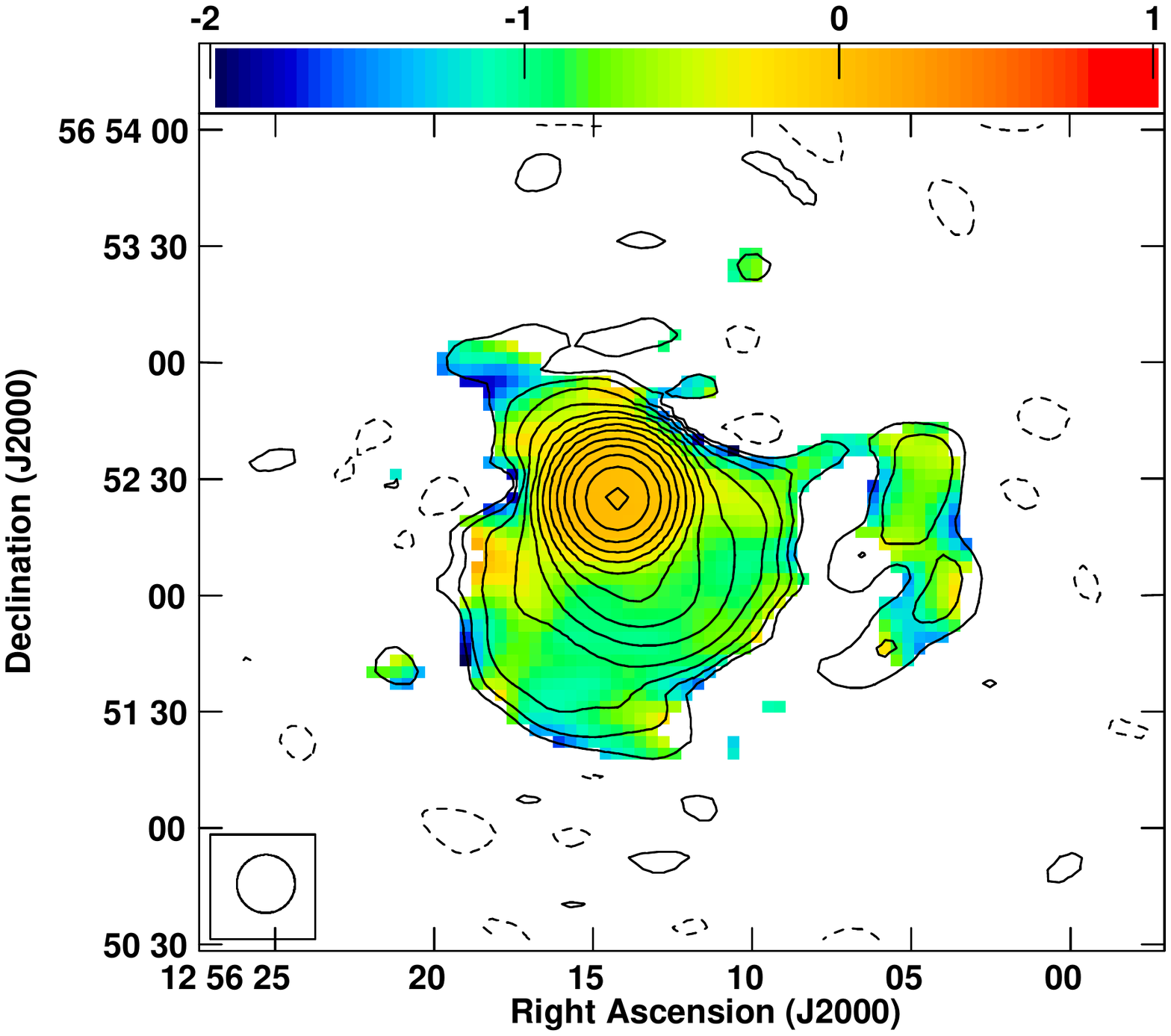}
\includegraphics[width=6.2cm,trim=60 120 90 160]{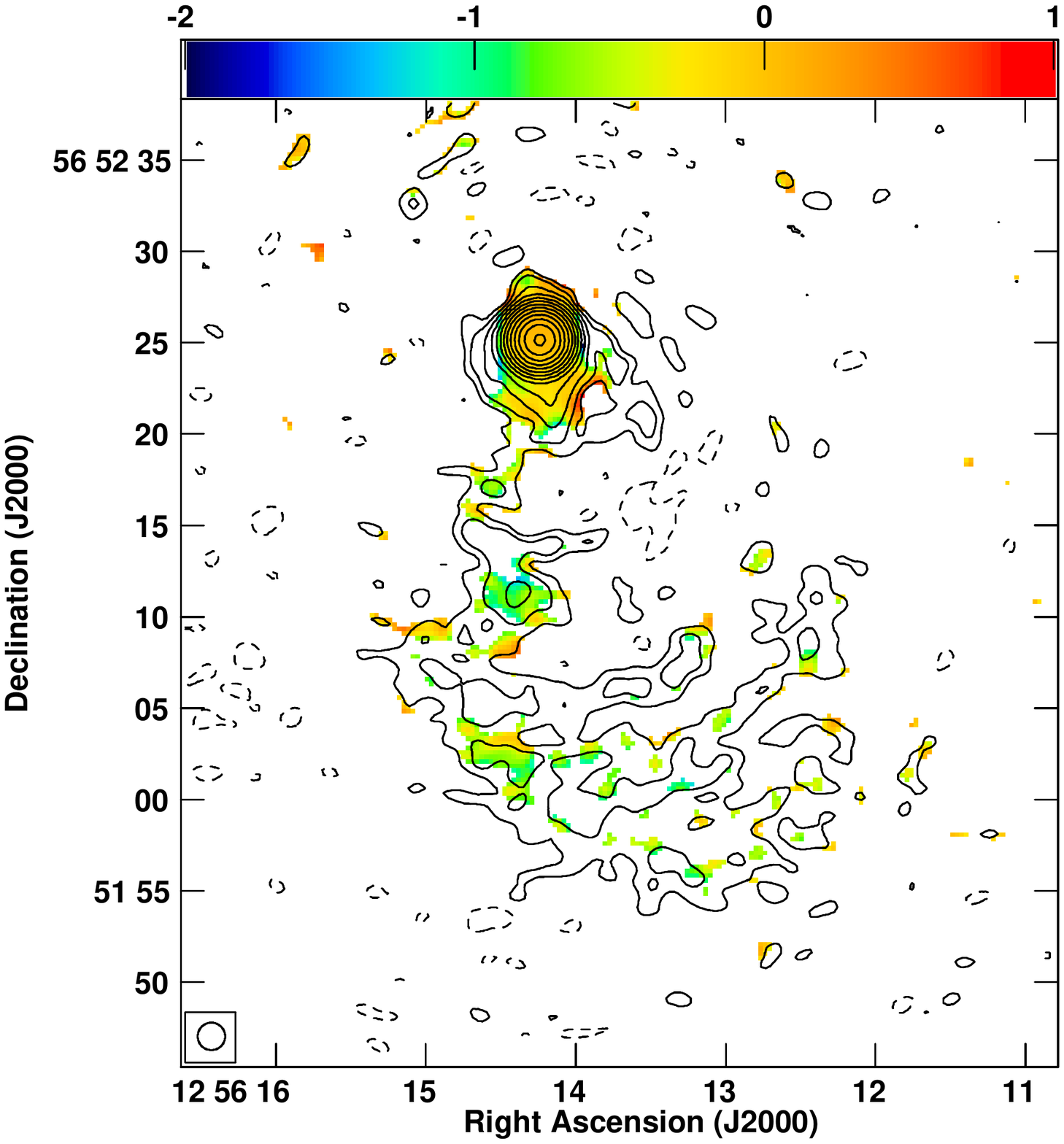}}
\caption{\small $1.43-4.86$ GHz spectral index images in colour with total intensity contours. Left: 15 arcsec resolution image. Contour levels are in percentage of peak surface brightness (=271.1~mJy~beam$^{-1}$), with contours levels increasing in steps of 2, with the lowest contour level being $\pm0.021$\%. The color scale extends from $-$2.0 to 1.0. Right: 1.5 arcsec resolution image. Contour levels are in percentage of peak surface brightness (=232.3~mJy~beam$^{-1}$), with contours levels increasing in steps of 2, with the lowest contour level being $\pm0.021$\%. The color scale extends from $-$2.0 to 1.0. }
\label{fig7}
\end{figure*}

Apart from the high levels of linear polarization and well-ordered B-fields indicating non-thermal jet emission, the  non-thermal nature is also confirmed by the $1.4-4.9$~GHz spectral index images presented in the left and right panels of Figure~\ref{fig7}. The $1.4-4.9$~GHz spectral index images reveal a flat spectrum radio core ($\alpha=-0.25 \pm 0.02$ for 15$\arcsec$ and $\alpha=-0.14 \pm 0.04$ for 1.5$\arcsec$ resolution images), consistent with the synchrotron self-absorbed base of a relativistic jet, and a steep spectrum lobe emission to the south ($\alpha=-0.81 \pm 0.09$ for 15$\arcsec$ and $\alpha=-0.5 \pm 0.4$ for 1.5$\arcsec$ resolution images), consistent with optically thin synchrotron emission.

\subsection{Wind}
\label{sec:4.2.2}
The rotating $\chi$ vectors at the core edges of Mrk\,231 (Figure~\ref{fig1}) suggest tangential or compressed B-fields, assuming optically thin emission. 
They could either be indicative of B-fields compressed along the edges by an expanding cocoon driven by a wind resulting from the disruption of an inclined jet \citep{Mukherjee18} or could represent the base of a magnetically-driven AGN wind \citep{Palumbo20} or a magnetized starburst-driven wind (see Section~\ref{sec4.1}) threaded by toroidal B-fields. Moreover, the sub-kpc scale disk-like emission detected in the 1.4 GHz VLBA image of \citet{Carilli98} \citep[see also][]{Morganti16} may represent the base of the wind.

The transverse B-fields revealed in the 1.4 GHz C-array and 4.9~GHz D-array images (Figure~\ref{fig3}) could be interpreted as arising from the ordering and amplification of B-fields by compression due to a series of transverse shocks, or could represent a toroidal component of a large-scale helical B-field associated with the jet \citep{Gabuzda94,Lister98,Pushkarev17}. However, the works of \citet{Miller12} and \citet{MehdipourCostantini19} suggest an association of toroidal B-fields with magnetically-driven AGN winds and poloidal B-fields with jets. Therefore, one cannot rule out the possibility of a magnetized accretion disk wind or the outer layers of a broadened jet \citep[like a jet sheath;][]{Mukherjee18} threaded by toroidal B-fields, sampled on larger spatial scales \citep[e.g., in III Zw 2;][]{Silpa21}.


\subsection{Jet \& Wind Composite}
\label{sec:4.2.3}
We find that the multi-scale polarization images presented in the current work probe different layers of the kpc-scale radio emission that are sampled at different resolutions. The parallel inferred B-fields in the regions close to the core could represent a poloidal B-field component in the spine of the weakly collimated jet. The transverse inferred B-fields in the core are suggestive of a toroidal B-field component at the base of the outflow, which continues all the way up to the edge of the southern lobe. The transverse inferred B-fields in the lobe could represent B-field amplification due to shock compression by a radiatively driven AGN wind, or toroidal B-fields threading a magnetized accretion disk wind or the outer layers of a broadened jet.

Thus, our results so far point to a composite jet and ``wind'' outflow in Mrk\,231, where the weakly collimated AGN jet with poloidal inferred B-fields is immersed inside a broader magnetized ``wind'' with toroidal inferred B-fields (see also Section~\ref{sec:4.2.4} that discusses the idea of composite jet and wind outflow in Mrk\,231, based on the energetic arguments). The ``wind'' may either comprise both nuclear starburst (close to the core) and AGN winds, or the outer layers of a broadened jet. Such a co-axial jet and wind outflow would appear co-spatial in projection, similar to that proposed in III~Zw~2 \citep{Silpa21}. Such a co-axial outflow comprising of jet and MHD disk wind, has also been revealed in the protostellar system HH 212 \citep{Lee21}.


\citet{Reynolds09} show the dereddened radio-loudness of Mrk\,231 to be $1.4<\mathrm{R}<3.8$, making it a radio-quiet AGN. Interestingly, \citet{Reynolds17} suggest that Mrk\,231 is transitioning from radio-quiet to radio-loud state. In this respect as well as other properties, Mrk\,231 shows close similarities to the radio-intermediate AGN III Zw 2 \citep{Brunthaler00,Silpa21}.

\subsection{Relative contributions of Jet \& Wind}
\label{sec:4.2.4}
In Section~\ref{sec:3.4.1}, we derive $\eta_\mathrm{jet} = 0.012$ using the relation by \citet{MerloniHeinz07}. This is in agreement with other measurements of $\eta_\mathrm{jet}$ in the literature \citep[for e.g.,][]{Balmaverde08, NemmenTchekhovskoy15}. The implications of $\eta_\mathrm{jet} \sim 0.01$ are: (1) about 1\% of the rest mass energy of accreted matter is extracted. (2) the mechanism for the extraction might be \citet{BlandfordZnajek77}.  
We note that $\eta_\mathrm{jet}$ derived using \citet{Willot99} and \citet{Punsly05} relations is lower than that derived using the \citet{MerloniHeinz07} relation by 2 orders of magnitude. This could suggest that the \citet{Willot99} model underestimates the fraction of the accreted rest mass energy which is carried away by the jet in radio-quiet AGN like Mrk\,231, given that \citet{MerloniHeinz07} relation is derived for a sample of low-luminosity AGN whereas \citet{Willot99} relation is derived for a sample of radio-powerful AGN. Moreover, it is well-known that the intrinsic flux density of a relativistic jet is Doppler-beamed in the direction towards the observer and at small viewing angles of the jet, the beamed emission appears even stronger \citep{CohenLister07, PearsonZensus87}. \citet{MerloniHeinz07} have corrected for the relativistic beaming effects in
their relation. This should take care of any potential disparity in the energetic analysis presented here, that may arise due to the outflow in Mrk\,231 being oriented close to the line of sight.

Interestingly, we obtain consistent values for S$_\mathrm{SBwind}$ using both \citet{Leitherer99} and \citet{DallaVecchiaSchaye08} models but different values for S$_\mathrm{AGNwind}$ (see Sections~\ref{sec:3.4.2} and \ref{sec:3.4.3}). The total radio luminosity estimated (say, S$_\mathrm{tot}$) is 294~mJy in 4.9~GHz D-array image, 308~mJy in the 1.4~GHz C-array image and 318~mJy in the 1.4~GHz D-array image. Therefore, our estimation suggests that the starburst-driven wind accounts for $\sim$ 8\% of the total radio emission detected in the lower resolution image at 4.9~GHz and $\sim$ 20\% of the total radio emission detected in the lower resolution image at 1.4~GHz in Mrk\,231. Using the estimates of S$_\mathrm{SBwind}$ and S$_\mathrm{AGNwind}$, we now estimate the contribution of the weakly collimated jet to the total radio emission in the lower resolution images. Essentially, we estimate: $\mathrm{S_{jet} = S_{tot}-(S_{SBwind}+S_{AGNwind})}$. We also note that the relative contributions of jet and wind will change with a change in the coupling efficiency.

For a coupling efficiency of 5\%, the \citet{Leitherer99} model yields S$_\mathrm{AGNwind}$ $\sim$ 360~mJy at 1.4~GHz and $\sim$ 130~mJy at 4.9~GHz. This gives S$_\mathrm{jet}$ = 294 $-$ (23+130) = 141 mJy in the 4.9~GHz D-array image. We note that the value of S$_\mathrm{AGNwind}$ obtained at 1.4~GHz is slightly higher than S$_\mathrm{tot}$ obtained in the 1.4 GHz C- and D-array images. On the other hand, the \citet{DallaVecchiaSchaye08} model yields S$_\mathrm{AGNwind}$ $\sim$ 1.1~Jy at 1.4~GHz and $\sim$ 400~mJy at 4.9~GHz assuming \citet{Chabrier03} IMF and $\sim$ 1.8~Jy at 1.4~GHz and $\sim$ 650~mJy at 4.9~GHz assuming \citet{Salpeter55} IMF. We note that the value of S$_\mathrm{AGNwind}$ obtained using \citet{DallaVecchiaSchaye08} model is higher than S$_\mathrm{tot}$ for all three images. This could suggest that the \citet{DallaVecchiaSchaye08} model over-estimates the AGN wind contribution and predicts negligible contributions from starburst-driven wind and radio jet for a source like Mrk\,231.

For a coupling efficiency of 0.5\%, the \citet{Leitherer99} model yields S$_\mathrm{AGNwind}$ $\sim$ 36~mJy at 1.4~GHz and $\sim$ 13~mJy at 4.9~GHz. This gives S$_\mathrm{jet}$ = 294 $-$ (23+13) = 258 mJy in the 4.9~GHz D-array image, S$_\mathrm{jet}$ = 308 $-$ (64+36) = 208 mJy in the 1.4~GHz C-array and S$_\mathrm{jet}$ = 318 $-$ (64+36) = 218 mJy in the 1.4~GHz D-array image. On the other hand, the \citet{DallaVecchiaSchaye08} model yields S$_\mathrm{AGNwind}$ $\sim$ 110~mJy at 1.4~GHz and $\sim$ 40~mJy at 4.9~GHz assuming \citet{Chabrier03} IMF and S$_\mathrm{AGNwind}$ $\sim$ 180~mJy at 1.4~GHz and $\sim$ 65~mJy at 4.9~GHz assuming \citet{Salpeter55} IMF. The \citet{Chabrier03} IMF yields S$_\mathrm{jet}$ = 294 $-$ (23+40) = 231 mJy in the 4.9~GHz D-array image, S$_\mathrm{jet}$ = 308 $-$ (64+110) = 134 mJy in the 1.4~GHz C-array and S$_\mathrm{jet}$ = 318 $-$ (64+110) = 144 mJy in the 1.4~GHz D-array image. The \citet{Salpeter55} IMF yields S$_\mathrm{jet}$ = 294 $-$ (23+65) = 206 mJy in the 4.9~GHz D-array image, S$_\mathrm{jet}$ = 308 $-$ (64+180) = 64 mJy in the 1.4~GHz C-array and S$_\mathrm{jet}$ = 318 $-$ (64+180) = 74 mJy in the 1.4~GHz D-array image.

\citet{Feruglio15} find that the energy outflow rates of the ultra-fast outflow and molecular outflow traced by CO in Mrk\,231 are in agreement with the coupling efficiency required in the \citet{DiMatteo05} AGN feedback model. \citet{Morganti16} find that the jet power is not sufficient to drive the HI and molecular outflows in Mrk\,231. These works favour a wide-angle wind origin for these gas outflows based on morphology and energetics. \citet{RupkeVeilleux11} find that the mass and energy outflow rates of the neutral atomic outflow traced by Na~I~D in Mrk\,231 are consistent with the coupling efficiency required in the \citet{HopkinsElvis10} AGN feedback model. 
We note that when 0.5\% coupling efficiency \citep[i.e., the][model]{HopkinsElvis10} is assumed, the radio contribution of the jet estimated using \citet{Leitherer99} model and \citet{DallaVecchiaSchaye08} model assuming \citet{Chabrier03} IMF, surpasses that of the wind. Overall, it thus appears that although the cold gas outflows favour a wide-angle wind origin, one cannot unambiguously rule out the presence of a low-power weakly collimated large-scale jet in Mrk\,231. 

Another interesting finding is that using S$_\mathrm{jet}$ estimated from the 4.9~GHz D-array image in equations (5)-(7), we obtain $\eta_\mathrm{jet}$ of $\sim$0.007 with 5\% coupling efficiency and $\sim$0.011 with 0.5\% coupling efficiency. 
Thus, $\eta_\mathrm{jet}$ estimated using higher and lower resolution images are broadly consistent with each other. Overall, the coupling efficiency can have a range of values without changing the overall picture in Mrk\,231. We find a broad agreement between the energetics of the radio outflows derived using the approach discussed in Section~\ref{sec3.4}, and the energetics of the neutral atomic and molecular outflows reported in the literature, suggesting that our approach is valid and close to reality.

\subsection{Curvature of the kpc-scale jet}
The curved kpc-scale weakly collimated jet in Mrk\,231 could arise either from an interaction between the jet and the surrounding medium or from a precessing jet. Jet precession could result from a warped accretion disk or a binary black hole \citep{Pringle96, CaproniAbraham04, Krause19} or a precessing accretion disk owing to disc instabilities from an ongoing merger event \citep{Liska18}. The disturbed morphology of the host galaxy and the presence of tidal tails suggest that Mrk\,231 is probably at the final stage of a merger \citep{Armus94, Lipari94}. 

A binary black hole has been proposed in Mrk\,231 by \citet{Yan15}, in order to explain the unusual optical-UV spectrum observed in this source. They suggest that while the smaller-mass black hole ($4.5\times10^6$~M$_{\sun}$) accretes as a thin accretion disk and the larger-mass black hole ($1.5\times10^8$~M$_{\sun}$) accretes via radiatively inefficiently advection dominated accretion flow (ADAF), the two black holes are also surrounded by a circumbinary disk. However, the binary black hole scenario has come under scrutiny by recent infrared and ultraviolet spectral studies of Mrk\,231 \citep[e.g.,][]{Veilleux16,Leighly16}. Hence, there is presently no strong observational evidence in favor of the binary black hole in Mrk\,231.

\subsection{Matter-dominated outflow}
\label{sec4.4}
A ``failed'' jet scenario has been proposed in Mrk\,231 \citep{RupkeVeilleux11,Morganti16,Reynolds20}. \citet{Wang21} argue that the jet in Mrk\,231 is obstructed by the dense ISM within a few tens pc scales, which prevents its growth to larger scales. A low ionization BAL wind \citep{Lipari94, Smith95}, a high ionization X-ray absorbing wind \citep{Feruglio15, Reynolds17} and huge amounts of circumnuclear dust \citep{Smith95}, all contribute towards a dense nuclear environment that may slow down or disrupt the jet. 

Sub-relativistic expansion of the pc-scale jet in Mrk\,231 \citep{Ulvestad99b} is similar to the `slow' phase seen in III Zw 2 \citep{Brunthaler00}. This suggests that the region of relativistic expansion in Mrk\,231 is on even smaller spatial scales, or that the expansion is intrinsically slow quite early on. The multi-scale radio morphology and spectral index values (Section 3.1), polarization structures (Section 3.2) and energetic arguments (Sections 3.5 and 4.4) are consistent with the presence of a kpc-scale weakly collimated jet in Mrk\,231. Therefore, it is possible that the jet in Mrk\,231 was launched relativistically but slowed down without getting completely disrupted.

As discussed in Section~\ref{sec3.3}, we conclude that for both poloidal inferred B-fields associated with the spine of the weakly collimated jet and toroidal inferred B-fields associated with the wind/jet sheath, $\beta$ increases with increasing distance from the core, implying that the composite jet and wind outflow in Mrk\,231 becomes increasingly matter-dominated away from the core \citep[e.g.,][]{Meier03}. Simulations by \citet{Mukherjee20} suggest that the matter-dominated and low-power jets are more susceptible to instabilities than the Poynting flux dominated jets, and therefore could get easily deaccelerated/decollimated but not terminated, such as that seen in
Mrk 231. 

\section{Summary and Conclusions}
Our multi-frequency, multi-scale polarization-sensitive VLA observations of the Seyfert 1 galaxy and BALQSO, Mrk\,231 at 1.4 and 4.9~GHz reveal the likely presence of multiple radio components with characteristic B-field geometries. We summarise our primary findings below. 

\begin{enumerate}
\item Mrk\,231 shows a core-dominated radio source with extended emission in total intensity on kpc-scales with the VLA. A combination of a weakly collimated jet and a wind or jet sheath, oriented at a small angle to line of sight, can explain the observed total intensity and polarization structures.

\item The ``wind'' component of the composite outflow may be driven by both a starburst (close to the core) and the AGN, where the latter maybe the primary driver. Moving away from the core, the ``wind'' component may also comprise the outer layers (or sheath) of a weakly collimated jet.

\item The model of a kpc-scale weakly collimated jet/lobe in Mrk\,231 is strengthened by its C-shaped morphology, steep spectral index throughout, complexities in the B-field structures, and the presence of self-similar structures observed on the 10-parsec-scale in the literature. The latter may even suggest the presence of episodic jet activity in Mrk\,231.

\item Our results suggest that the starburst-driven wind accounts for $\sim$ 8\% of the total radio emission detected in the lower resolution image at 4.9~GHz and $\sim$ 20\% of the total radio emission detected in the lower resolution image at 1.4~GHz in Mrk\,231. 

\item The inferred value of the (weakly collimated) jet production efficiency, $\eta_\mathrm{jet} \sim 0.01$ using the relation by \citet{MerloniHeinz07}, is consistent with other estimates in the literature and its implications are: (1) about 1\% of the rest mass energy of accreted matter is extracted. (2) the mechanism for the extraction might be \citet{BlandfordZnajek77}. 

\item Overall, we conclude that kpc-scale weakly collimated radio jet in Mrk\,231 with poloidal inferred B-fields is curved and one-sided, and immersed inside a broader magnetized wind with toroidal inferred B-fields, with the composite outflow being matter-dominated.
\end{enumerate}

\section*{Acknowledgements}
We thank the referee for their insightful comments. The National Radio Astronomy Observatory is a facility of the National Science Foundation operated under cooperative agreement by Associated Universities, Inc. SB and CO are partially supported by the  Natural Sciences and Engineering Research Council (NSERC) of Canada. We acknowledge the support of the Department of Atomic Energy, Government of India, under the project 12-R\&D-TFR-5.02-0700.\\

\section*{Data Availability}
The data underlying this article will be shared on reasonable request to the corresponding author. The VLA data underlying this article can be obtained from the NRAO Science Data Archive (https://archive.nrao.edu/archive/advquery.jsp) using the proposal id: AB740. 



\bibliographystyle{mnras}
\bibliography{mnras} 






\bsp	
\label{lastpage}
\end{document}